\documentclass[a4paper,10pt]{article}

\usepackage{joneseystyle}
\usepackage{geometry}
\usepackage{float,textcomp}
\usepackage{booktabs,multirow}
\usepackage[font={footnotesize,up},skip=0pt]{caption}
\usepackage[margin=2ex,font=scriptsize]{subcaption}
\usepackage{wrapfig}
\usepackage[pdftex]{graphicx}
\usepackage{amssymb}
\usepackage{cite}
\usepackage[cmex10]{amsmath}
\usepackage{amsthm}
\usepackage{dsfont}

\usepackage{color}

\begin{document}

\title{Continuous Compressed Sensing of Inelastic and Quasielastic Helium Atom Scattering Spectra}
\author{Alex Jones \\ DAMTP \\ University of Cambridge
 \and
Anton Tamt{\"o}gl \\ Cavendish Laboratory \\ University of Cambridge
\and
Irene Calvo-Almaz{\'a}n \\ Cavendish Laboratory \\ University of Cambridge
 \and
Anders Hansen \\ DAMTP \\ University of Cambridge }

\maketitle

\begin{abstract}
Helium atom scattering (HAS) is a well established technique for examining the surface structure and dynamics of materials at atomic sized resolution. The HAS technique Helium spin-echo spectroscopy opens up the possibility of compressing the data acquisition process. Compressed sensing (CS) methods demonstrating the compressibility of spin-echo spectra are presented. In addition, wavelet based CS approximations, founded on a new continuous CS approach, are used to construct continuous spectra that are compatible with variable transformations to the energy/momentum transfer domain. Moreover, recent developments on structured multilevel sampling that are empirically and theoretically shown to substantially improve upon the state of the art CS techniques are implemented. These techniques are demonstrated on several examples including phonon spectra from a gold surface.
\end{abstract}

\section{Introduction}

Helium Spin-Echo (HeSE) spectroscopy is the ideal tool for studying the dynamical behaviour of a wide variety of surfaces ranging from simple metals to reactive and metastable surfaces\cite{He3Apparatus}. HeSE provides information on the static structure of a surface in a manner analogous to Bragg's original work \cite{Bragg} on X-ray crystallography, but more importantly it can be used to study the dynamics on atomic length scales with picosecond time resolution\cite{He3Apparatus}. This involves atoms and molecules diffusing on the surface \cite{ohninereview}, phonon vibrations \cite{oldphononpaper}, etc. Thanks to the work of van Hove \cite{vanHove} et al. a theoretical framework exists which describes the dynamics of atoms and molecules through the relation of time and position $(\mb{R},t)$ to momentum and energy transfer $(\Delta\mb{K},\Delta E)$ as a Fourier pair.

In this paper we present a Compressed Sensing (CS) approach for compressing this measurement process, showing that the time needed to reconstruct HeSE spectra can be reduced by several orders of magnitude compared to standard Discrete Fourier Transform (DFT) reconstruction techniques. CS, pioneered by Cand{\`e}s, Donoho, Tao et. al. \cite{CandesTao,Donoho} has long been associated with Nuclear Magnetic Resonance (NMR) based applications such as Magnetic Resonance Imaging \cite{CompressedMRI, Unser} and NMR Spectroscopy \cite{CompressedSpectroscopy}. Recently, compressed sensing has also seen applications focusing on Raman spectroscopy measurements \cite{RamenApplication} and in molecular dynamics simulations \cite{MDApplication}.

\begin{figure}[t]
\centering
\includegraphics[width=0.80\textwidth]{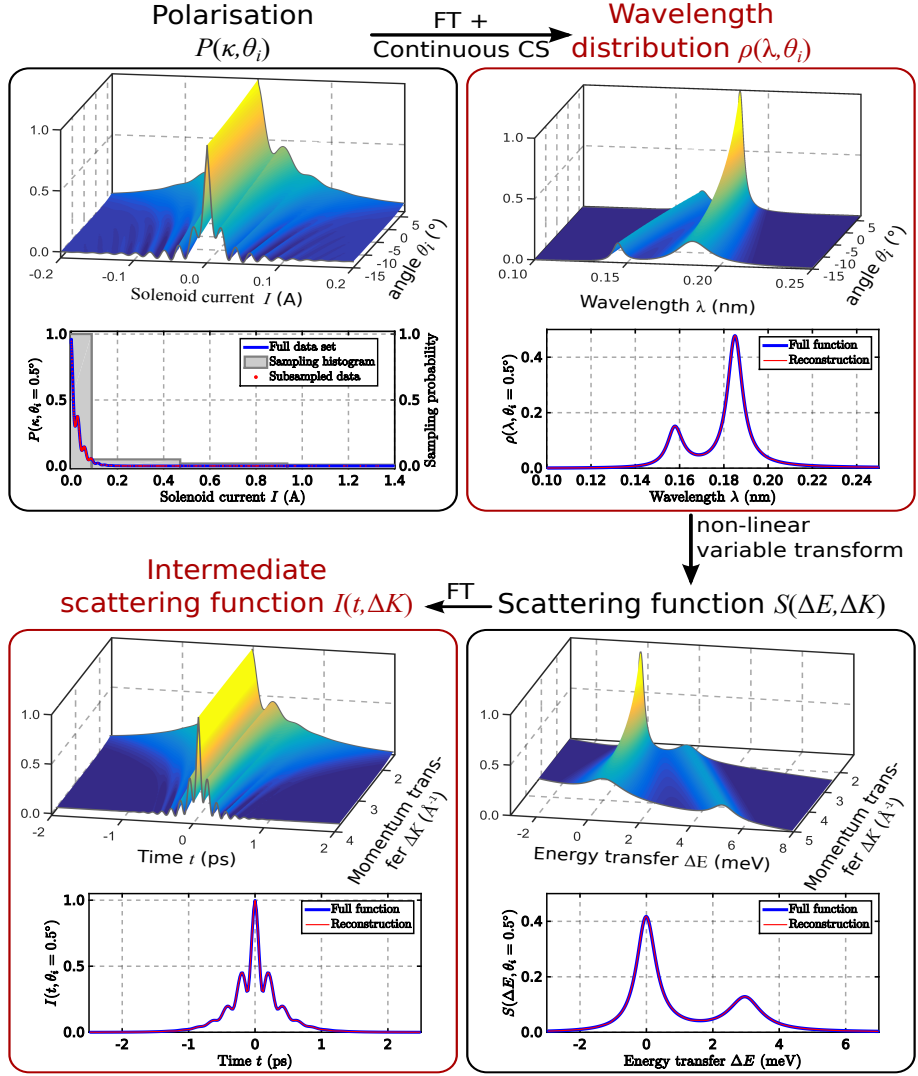}

\vspace{10pt}

\caption{Diagram outlining the various stages of data transformation from measurement in polarisation to the Intermediate Scattering Function (ISF) describing molecular processes. The Upper plots denote the full 2D data while the lower plots are 1D projections/slices. The stages highlighted in red correspond to the target data we wish to reconstruct. The plotted intensities are in consistent arbitrary units (a.u.). Current spacing / resolution is $2.7 \cdot 10^{-4}$A / 10233 points and 930 points ($\approx 9\%$) are subsampled. The experimental variable $\kappa$ is proportional to current $I$ (see \eqref{Eq:Current2Kappa}).}
\label{Fig:Figure1_4.eps}
\end{figure}

Spin-echo spectroscopy shares clear similarities with these fields, such as Fourier transforms arriving naturally in data acquisition, however there are also significant differences. In particular, one of the goals of spin-echo spectroscopy is to determine dynamical processes by monitoring the change of polarisation data. Here we consider the whole process of data processing, from polarisation data measurements to the extraction of the molecular dynamics information. Unlike NMR and many other spectroscopy-based applications, after we have performed compression on the initial Fourier transform step we cannot directly use the output data, it must instead undergo several further transforms. Crucially, this includes a non-linear change of variables to momentum/energy space $(\Delta\mb{K},\Delta E)$. This precludes the use of standard DFT-based CS techniques as this transform distorts the (necessarily discrete) set of values that we can solve for the wavelength intensity function.

Instead a new \emph{continuous} CS approach, recently introduced by Adcock, Hansen et. al \cite{ContinuousCompressed,LeveledCompressed,AHBogdanTeshckeGenSamp}, is used to reconstruct a continuous approximation that avoids discretising the wavelength intensity function entirely. With this method one has the freedom of evaluating the reconstructed wavelength intensity function at any point they desire while still having the speed-up benefits of compressive sampling. Such an approach could be used to handle other inverse problems that require further transformations after reconstruction. Moreover, as demonstrated in \cite{ContinuousCompressed,LeveledCompressed,AHBogdanTeshckeGenSamp} the continuous model avoids the so-called wavelet and inverse ``crimes'' and hence provides superior reconstructions compared to the classical approach.

Figure 1 shows both the process of converting sampled polarisation data to the intermediate scattering function alongside the application of this new CS technique. Note that CS reconstructions in red for the wavelength intensity and scattering functions match the true signal with $9\%$ of the data traditionally used to reconstruct such spectra using direct Fourier inversion without compression.

Moreover, in NMR-based experiments the smallest group of data that is taken in one measurement is typically a line (or path) of data points in k-space, however in HeSE each measurement corresponds to a single point. This gives us an additional degree of  freedom in the data acquisition process which makes this application of CS particularly effective, as one can utilize the new approach of structured multilevel sampling in \cite{LeveledCompressed} to its fullest to boost performance. In particular, by also taking the structure of the signal into account when designing the sampling strategy one can outperform the classical compressed sensing results (see \cite{Roman, WaveletSampling2} for experimental validation) that are dictated by the estimate on the number of samples $m$ to be
\[
m \gtrsim s \log(N),
\]
where $s$ is the number of non-zero or important coefficients and $N$ is the dimension of the vector (The notation $f(\cdot) \gtrsim g(\cdot)$ means there is a universal constant $C>0$ such that $f(\cdot) \ge C \cdot  g(\cdot)$).

 However, if a signal has $s = M_1 + s_2 + \hdots + s_r$ non-zero coefficients where $M_1$ denotes the number of the first consecutive non-zero coefficients in the first levels of a wavelet expansion and  $s_j$ is the number of non-zero coefficients in the $j$-th level of the wavelet structure then, by using a multilevel sampling procedure \cite{LeveledCompressed}, one needs only
$$
m \gtrsim M_1 + (s_2 + s_3 + \hdots + s_r) \log(N).
$$
measurements \cite{Foundations, LeveledCompressed}. Typically, the coefficients corresponding to $M_1$ are the most important, and most of the energy in the signal is contained in these. This is very convenient as we do not have to pay a log factor for these coefficients.  In practice this means substantial gain over the standard approaches as demonstrated in \cite{Roman, WaveletSampling2}.

\subsection{Compressing Spin-Echo Spectra}

To understand how continuous CS differs from conventional DFT CS we start with a typical 1D Fourier problem where we measure Fourier samples $P$ of a wavelength intensity function $\rho$ we want to reconstruct:

\begin{equation} \label{Eq:OneDimFourier}
P(\kappa)= \int \rho(\lambda) e^{2 \pi \ri \kappa \cdot \lambda} \, \D \lambda, \qquad \lambda , \kappa \in \bbR.
\end{equation}

From the above it is immediately recognised that $P$ is a Fourier Transform of $\rho$ and therefore $\rho$ can be obtained by the inverse Fourier transform of $P$. For a general function $\rho$ this would require knowing $P(\kappa)$ at every point $\kappa \in \bbR$ which is unrealistic.

In practice however, the wavelength intensity function $\rho$ is treated as a periodic function
\begin{equation} \label{Eq:PeriodicWavelength}
\tilde{\rho}(\lambda)= \sum_{k \in \bbZ} \rho(\lambda+(b-a)k),
\end{equation}
over a fixed interval $[a,b]$. This is convenient because it permits changing the problem to one of handling a Fourier transform to that of handling a Fourier series expansion:

\begin{equation} \label{Eq:FourierSeriesReduction}
\begin{aligned}
&\tilde{\rho}(\lambda) = \sum_{k \in \bbZ} \langle \rho, \chi_{k, \epsilon} \rangle \chi_{k, \epsilon}(\lambda), \quad \lambda \in [a,b],
\\
&\langle \rho, \chi_{k, \epsilon} \rangle= \int_{\bbR} \rho(\lambda) \overline{ \chi_{k, \epsilon}(\lambda)} \, d \lambda =P(k \epsilon) \epsilon.
\\
\chi_{k, \epsilon}(\lambda)& = \epsilon^{1/2} \exp(-2 \pi \ri \epsilon k \cdot  \lambda), \quad k \in \bbZ, \ \epsilon=(b-a)^{-1},
\end{aligned}
\end{equation}

The upshot of \eqref{Eq:FourierSeriesReduction} is that we now only need the values $P(k \epsilon), k \in \bbZ$, to obtain the wavelength intensity function $\rho$, rather than $P(\kappa), \kappa \in \bbR$.

Typically the next step is to truncate the Fourier series expansion, meaning that one makes the approximation
\begin{equation} \label{Eq:FourierTruncation}
\tilde{\rho}(\lambda) \approx \tilde{\rho}_N(\lambda) = \sum_{k =-N}^N  \langle \rho, \chi_{k, \epsilon} \rangle  \chi_{k, \epsilon}(\lambda), \quad \lambda \in [a,b],
\end{equation}
for some fixed $N \in \bbN$. The problem is now feasible as only finitely many data points $k$ are required to determine $\tilde{\rho}_N$.

Up to this point both continuous CS and conventional DFT CS agree. Conventional DFT CS then breaks up the interval $[a,b]$ into a uniform grid of $N$ points
\begin{equation} \label{Eq:DFTSimplifcation}
\lambda_{j,N}= a + \frac{b-a}{2N+1}j
\end{equation}
and solves for $\tilde{\rho}_N(\lambda_{j,N}), \ j=1,...,2N+1$. The advantage of doing this is that \eqref{Eq:FourierTruncation} becomes a vector-matrix equation of the form $g=Af$ where $g_j=\tilde{\rho}_N(\lambda_{j,N})$, $A$ is a DFT matrix and $f$ corresponds to samples of $P$. This can be inverted to give $f=A^{-1} g$ where $A^{-1}$ is still a DFT matrix, and therefore an isometry, which facilitates the application of CS.

The drawback of this approach is that we have lost something by only considering $\tilde{\rho}_N(\lambda_{j,N}), \ j=1,...,2N+1$ instead of \eqref{Eq:FourierTruncation}. The Fourier series approximation could have been evaluated at any point in the interval $[a,b]$ and suddenly the best we can do is reconstruct the $\tilde{\rho}_N(\lambda_{j,N})$, even though we are still working with the same number of Fourier samples. Do we really have to pay this price in order to be able to compress this problem?

Motivated by the Fourier series \eqref{Eq:FourierTruncation}, one can try approximating $\rho$ in terms of a new \emph{Reconstruction Basis} $\sigma_n, \ n \in \bbN$
\begin{equation} \label{Eq:FunctionApproximation}
\tilde{\rho}(\lambda) \approx   \sum_{n =1}^M \langle \rho , \sigma_n \rangle \sigma_n(\lambda), \quad \lambda \in [a,b].
\end{equation}
Apart from the benefit of the keeping the problem continuous, one also has the freedom to \emph{choose} which basis $\sigma_n$ to work with, making the approach more versatile than a straight DFT approach.

Since one is still sampling data that corresponds to Fourier coefficients of $\rho$, it is impossible to exclusively work with their choice of basis $\sigma_n$. Instead one has to convert Fourier series coefficients into coefficients in the basis $\sigma_n$. This is achieved by working with the infinite change of basis matrix for the two bases:

\begin{equation} \label{Eq:ContinuousChangeOfBasis}
B_{k,n}= \langle \sigma_n, \chi_{k, \epsilon} \rangle, \quad n \in \bbN, \ k \in \bbZ.
\end{equation}

Using this matrix to reconstruct the coefficients $\langle \rho , \sigma_n \rangle, n=1,...,M$ we can then use \eqref{Eq:FunctionApproximation} to approximate $\tilde{\rho}(\lambda)$. As long as we assume the $\sigma_n$ form an orthonormal basis then the matrix $B$ is an infinite-dimensional isometry which also allows the application of CS thanks to the work of \cite{ContinuousCompressed,LeveledCompressed, AHBogdanTeshckeGenSamp}.

This continuous approach has two significant advantages:
\begin{itemize}
\item The approximation $\tilde \rho$ is now a continuous function (as opposed to discrete) that can be evaluated at any point and hence this allows the non-linear change of variables going from the wavelength distribution to the scattering function $S$ (as shown in Figure 1). Such a transform is not possible with conventional discrete CS techniques.
\item The approximation $\tilde \rho$ is computed with the actual coefficients in the new expansion of the wavelength function, and hence this approximation has the characteristics of the approximation in the new basis rather than the truncated Fourier series. This means reducing Gibbs ringing and other artefacts coming from Fourier approximations (see \cite{ContinuousCompressed,LeveledCompressed,AHBogdanTeshckeGenSamp} for details).
\end{itemize}

Details on the theoretical background of CS, what basis to use, the convex optimisation problems we solve and how to subsample the Fourier data are provided later on in the paper.

\subsection{Paper Outline}

The rest of the paper is split into two halves. The first focuses on the first Fourier transform step from polarisation data to the wavelength intensity function shown in Figure 1. Focusing on the step, DFT CS is used to demonstrate the compressibility of phonon detection.

The second half goes into the full cycle of transforms shown in Figure 1, from polarisation data to molecular processes. After covering some brief background on molecular diffusion we elaborate on the continuous CS technique leading to a final comparison with DFT CS.

\section{Helium Spin-Echo Spectroscopy} \label{Sec:BackGrdHAS}

Surface-related phenomena are prevalent throughout everyday life through processes such as friction, corrosion and tension. When designing new materials, such as for catalysis \cite{TransportAdsorbates} or electronics \cite{ThermalCarbonNanotubes, AtomicWaterWheels}, one would like to study various vibrational and electronic properties to determine their potential benefits and drawbacks. Neutrons and X-rays have been successful in measuring such phenomena over the entire bulk of a sample. However, to truly understand surface processes one needs to investigate the differences between the structure of a surface and the bulk \cite{SurfacePhysicsIntro}. Therefore, a purely surface sensitive approach is highly desirable. The repulsive part of the $^3$He/surface interaction potential prevents the He atoms from penetrating into the surface layer of materials. Consequently they are ideal as a surface scattering probe.

The basic setup of the Helium spin-echo apparatus at Cambridge's Cavendish laboratory is given in Figure 2. A beam of thermal $^3$He is generated from the source in a fixed direction. The nuclear spins are polarised and then rotated by a solenoid before being scattering upon the target crystal surface. Afterwards any scattered He atoms heading in the direction of the detector are then rotated by a second solenoid and passed through another polarisation filter. Thereby the apparatus achieves an energy resolution of 3 \textmu eV and dynamical processes within a time window spanning from the sub-picosecond regime up to hundreds of picoseconds can be observed. A schematic for the apparatus can be found in \cite{He3Apparatus}.

Key variables that the operator can freely adjust include

\begin{itemize}
\item The currents $I_i, I_f$ that run through Solenoids A and B respectively
\item The scattering geometry, namely the angle of the surface normal relative to the source/detector setup
\end{itemize}


The incoming monochromatic He beam can be viewed as a plane wave with propagation wavevector $\mathbf{k} \in \bbR^3$ and angular frequency $\omega$:

\begin{equation} \label{Eq:PlaneWave}
\psi(\mb{r},t)=\exp(\ri (\mb{k} \cdot \mb{r}-\omega t)), \qquad \mb{r} \in \bbR^3, \quad t \in \bbR.
\end{equation}
Here, $\mb{r}$ denotes position and $t$ is time. The wavevector $\mb{k}$ and wavelength $\lambda$ are related to particle's momentum $\mb{p}$ by the de Broglie relations
\begin{equation} \label{Eq:BroglieRelation}
\mb{p}=\hbar \mb{k}, \quad |\mb{p}|=p=\frac{2 \pi}{\lambda}.
\end{equation}
Furthermore, the frequency $\omega$ is related to the particle energy $E$ by the relation
\begin{equation} \label{Eq:FrequencyEnergyRelation}
E=\hbar \omega.
\end{equation}

Using these relations we can treat $\mb{k}$ as representing momentum and $\omega$ as energy. Notice that by Formula \eqref{Eq:PlaneWave} we have identified two Fourier pairs $(\mb{k},\mb{r}), (\omega,t)$. However, at present these pairs only relate to the beam of Helium and not the crystal surface that it will be scattering upon.

\begin{figure}[t]
\centering
\includegraphics[width=0.82\textwidth]{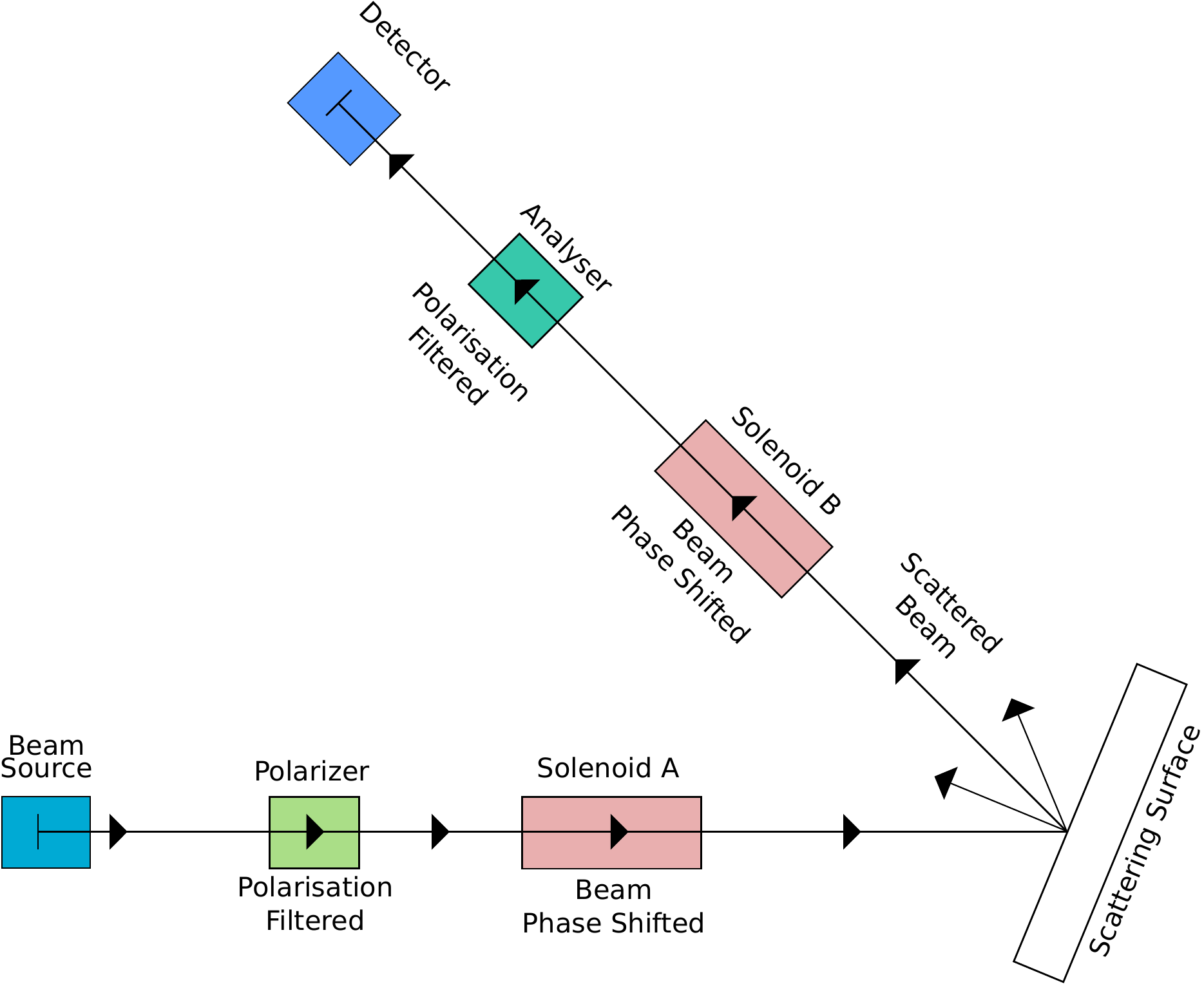}
\caption{Outline of the Cambridge Spin-Echo Apparatus.}
\label{Fig:BasicApparatus}
\end{figure}

\subsection{Phonons}

\begin{figure}[t]
\centering
\includegraphics[width=0.95\textwidth]{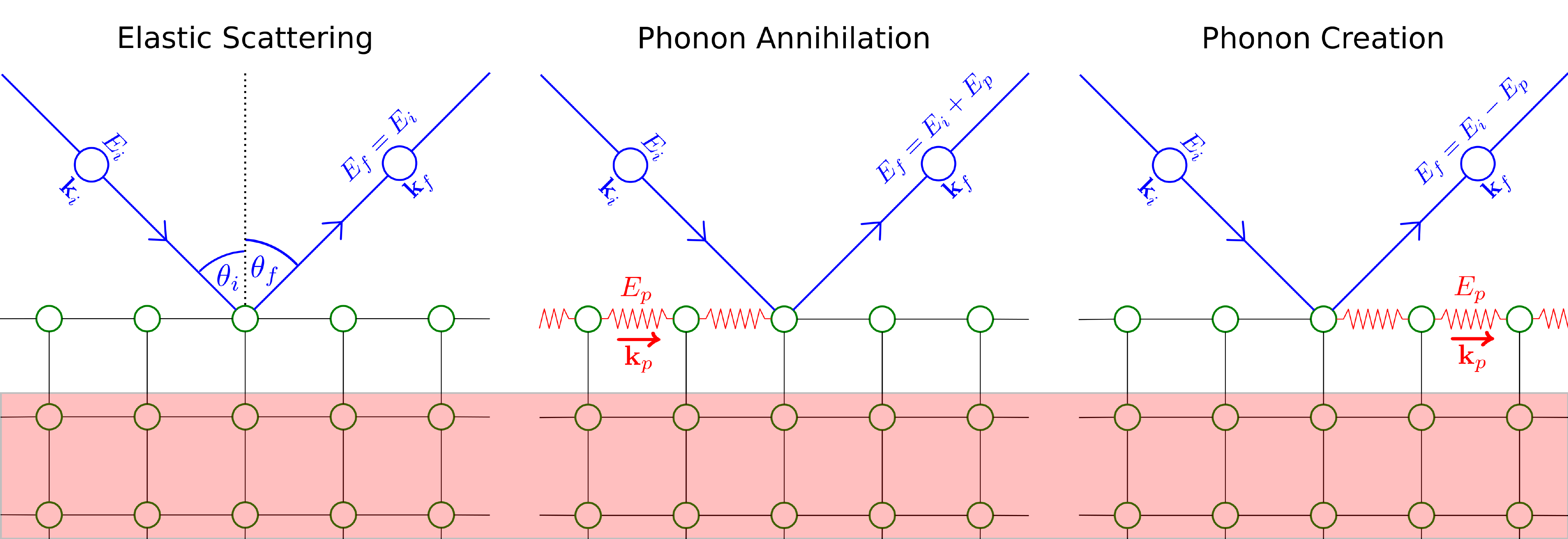}

\vspace{10pt}

\caption{Three possible scattering events. The $^3$He atom (in blue) can gain or lose energy after scattering. The $^3$He atom can only interact with the surface layer and therefore it can only interact with phonons on the surface. The He atom cannot penetrate into the bulk which is indicated by the red region.}
\label{Fig:PhononCreationAnnihilation}
\end{figure}

Detailed descriptions about lattice vibrations in a solid can be found in most solid state physics textbooks\cite{SolidStateIntro}.
At the surface the altered environment with respect to the bulk modifies the dynamics to give rise to new vibrational modes. This new vibrational modes are called surface phonons due to their localisation at the surface\cite{Benedek2010,Kress1991,Heid2003}.

Now we look at how phonons can influence the scattering of the He beam. Figure 3 shows three possible outcomes for the scattering of a single He atom on a crystal surface. The figure shows a cross section of the crystal where the path of the He beam lies in the same cross section, therefore, we are effectively working with a two dimensional problem. We shall assume there is conservation of energy before and after scattering and that the momentum is conserved parallel to the surface of the crystal (recall that the $^3$He beam cannot penetrate into the bulk which is why we only consider momentum parallel to the surface). If we let little $i, f$ denote the incident and final He states, little $p$ a possible phonon, $\mb{k}$ a wavevector and $k=|\mb{k}|$ its magnitude, then by conservation of energy and momentum
\begin{equation} \label{Eq:Conservation}
E_f=E_i+ \Delta E, \qquad K_i=K_f + \Delta K,
\end{equation}
where $K_i=k_i \sin \theta_i, K_f= k_f \sin \theta_f$ denotes the projections of the incident and final momentums $\mb{k}_i, \mb{k}_f$ onto the surface, provided that the scattering plane defined by $\mb{k}_i, \mb{k}_f$ contains the surface's normal vector. The change in energy $\Delta E$ is equal to the change in kinetic energy of the Helium particle:
\begin{equation} \label{Eq:KineticChange}
\Delta E = \frac{\hbar^2}{2m} k_f^2-\frac{\hbar^2}{2m} k_i^2.
\end{equation}
In the case of elastic scattering ($\Delta E =0$) we have $k_i=k_f$ and $\Delta K = G$ where $G$ corresponds to a reciprocal lattice vector consistent with the Laue equations. Suppose now that a He atom annihilates a surface phonon with energy $\Delta E = \hbar \omega_p$. For the momentum transfer $\Delta K = k_p + G$ holds where $\mb{k}_p$ is the wavevector of the phonon and $\mb{G}$ a surface reciprocal lattice vector and the norm of these vectors can be directly added if $\mb{\Delta K}$ is parallel to a high symmetry direction of the crystal.

The key point of these concepts is that the changes of energy and momentum contain information about the phonons on the crystal surface.

\subsection{Solenoid Currents and Spin Polarisation} \label{SubSec:CurrentSpin}

We now turn to how the scattering apparatus shown in Figure 2 can be used to measure properties of the He beam.  In particular, we show how the solenoid currents $(I_i,I_f)$ and scattering wavelengths $(\lambda_i,\lambda_f)$ share a direct Fourier relationship. This section follows closely the review \cite{GilJardine2D} of Alexandrowicz and Jardine and a more detailed description can be found in the Supporting Information.

Recall that we have two solenoids that generate magnetic fields which rotate the polarisation of the He beam. The solenoid current determines the strength of the magnetic field but it is more convenient to use the experimentally controllable parameter $\kappa$ which is proportional to the current in the solenoids via:
\begin{equation} \label{Eq:Current2Kappa}
\kappa_i= \frac{m \gamma B_{\mrm{eff}}  I_i}{2 \pi h}, \quad \kappa_f= \frac{m \gamma B_{\mrm{eff}}  I_f}{2 \pi h}.
\end{equation}
where $\gamma$ is the gyromagnetic ratio of the He atom and $B_{\mrm{eff}}$ is an apparatus specific constant. The polarisation of the He beam in terms of amplitude and phase can be conveniently written as a complex number. When using the scaled variables of equation \ref{Eq:Current2Kappa} the measured polarisation of the He beam in the detector
can be represented as the two-dimensional Fourier transform of $\rho(\lambda_i, \lambda_f)$:
\begin{equation} \label{Eq:TotalSignalWavelength2piVer}
\begin{aligned}
&P(\pmb{\kappa})  = \int \rho(\pmb{\lambda}) e^{2 \pi \ri \pmb{\kappa} \cdot \pmb{\lambda}} \, \D \pmb{\lambda}, \\
 \pmb{\lambda}&=(\lambda_i,\lambda_f) , \pmb{\kappa}=(\kappa_i,\kappa_f) \in \bbR^2.
\end{aligned}
\end{equation}
Here $\rho(\lambda_i, \lambda_f)$ denotes the \emph{Wavelength Intensity Function} describing the distribution of He atoms that reach the detector according to initial and final wavelengths.

\begin{figure}[t]
\centering
\includegraphics[width=0.82\textwidth]{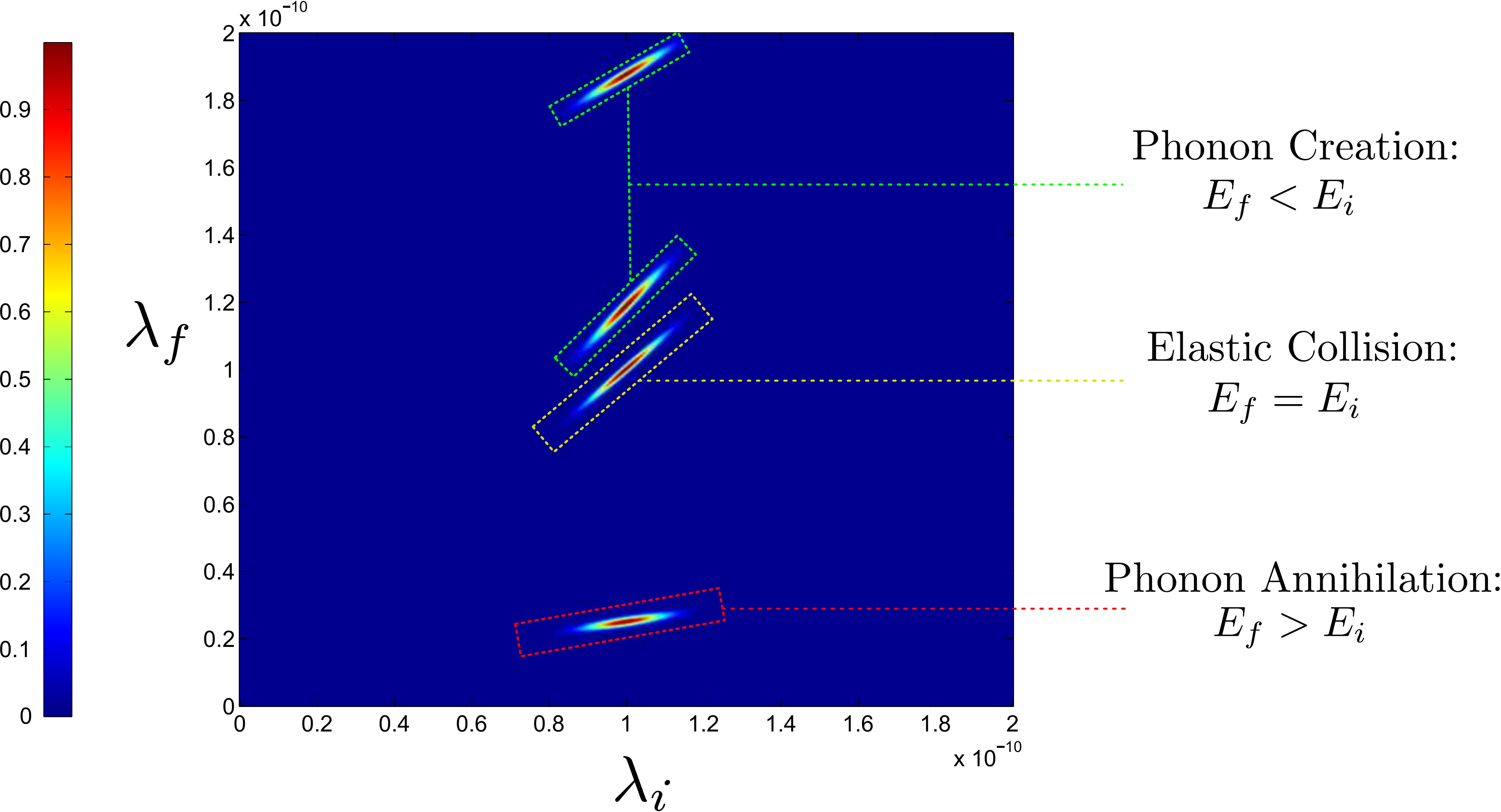}
\caption{Example of a wavelength intensity function with key features highlighted. Classification of the features is deduced from Equation \eqref{Eq:KineticChangeWavelength}.}
\label{Fig:WavelengthIntensityFunction}
\end{figure}
An example of a wavelength intensity function is given in Figure 4, along with possible phonon creation and annihilation events labelled. Another 2D example can be found in \cite{2DRecent} for incoming/outgoing energy spectra. Assuming the features in the plot originate from phonon phenomena on the crystal surface, the classification into creation/annihilation/elastic originates from Equation \eqref{Eq:KineticChange}:
\begin{equation} \label{Eq:KineticChangeWavelength}
E_f - E_i = \frac{\hbar^2}{2m} \Big( k_f^2-k_i^2 \Big) = \frac{\hbar^2}{2m} \Bigg( \frac{1}{\lambda_f^2}-\frac{1}{\lambda_i^2} \Bigg).
\end{equation}

\subsection{The Fourier Slice Theorem} \label{SubSec:FourierSlice}

Since the wavelength intensity function is supported around the average initial wavelength it could be viewed as a one-dimensional function that has been smoothed out into two-dimensions by the spread of initial wavelengths. Looking at Figure 4 as an example, we see that the key features of the wavelength intensity function can be broken down into lines of various slants which suggests that treating the function as one-dimensional would be advantageous.

It is precisely because of this decomposition into slanted regions that the two-dimensional problem is often reduced to a one-dimensional one using the Fourier slice theorem: Suppose we rotate the $\pmb{\lambda}=(\lambda_i, \lambda_f)$ coordinate system by an angle $\alpha$ to a new system $\pmb{\tau}=(\tau_1,\tau_2)=R_\alpha (\pmb{\lambda})$. Then one can derive the formula
\begin{equation} \label{Eq:PolarisationFormula}
\begin{aligned}
P_\alpha(\kappa_i)&:=P(R_{\alpha}(\kappa_i,0))=P(\kappa_i \cos \alpha, -\kappa_i \sin \alpha)
 \\ & = \int \Big( \int \rho(R_{\alpha} (\tau_1,\tau_2))  \, \D \tau_2 \Big) \exp(2 \pi \ri \kappa_i \tau_1) \, \D \tau_1.
\end{aligned}
\end{equation}
\begin{figure}[t]
\centering
\includegraphics[width=0.95\textwidth]{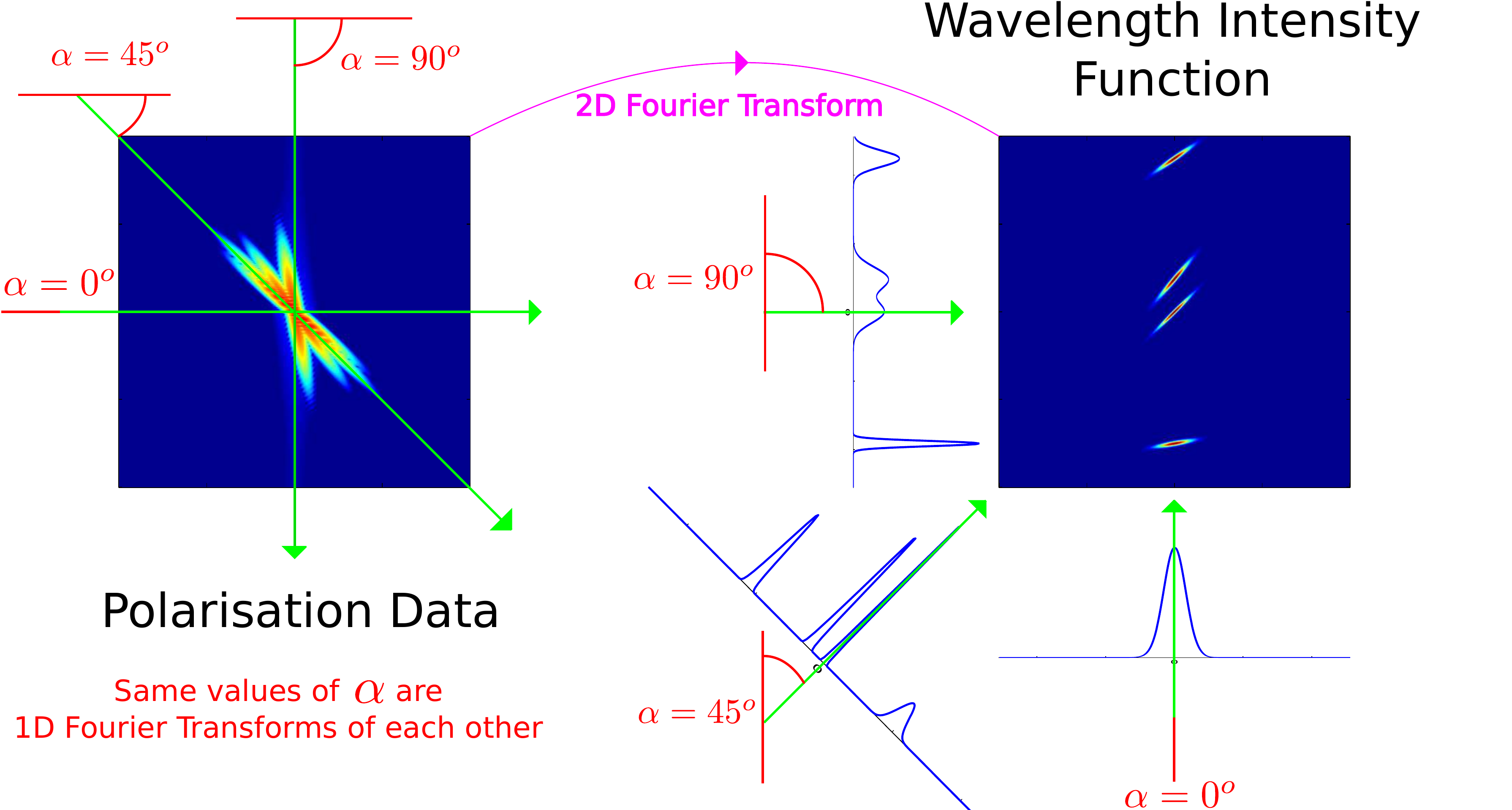}
\caption{Demonstration of the Fourier slice phenomenon. On the left we have the polarisation data and on the right we have the (approximation to) the wavelength intensity function. On the right the green lines indicate the direction of integration and the corresponding projection is shown as graph in blue.  The green lines on the left hand side represent the one-dimensional Fourier transforms of the projections shown on the right. The polarisation data intensity is shown on a log scale for the sake of readability.}
\label{Fig:FourierSliceExample}
\end{figure}
Therefore if we let $\rho_\alpha(\tau_1)$ denote the integral of $\rho(\lambda)$ along the line $\{R_{\alpha} (\tau_1,\tau_2) : \tau_2 \in \mathbb{R} \}$ then we have
\begin{equation} \label{Eq:PolarisationFormula2}
P(\kappa_i \cos \alpha,-\kappa_i \sin \alpha)= \int \rho_{\alpha}(\tau_1)\exp(2 \pi \ri \kappa_i \tau_1) \, \D \tau_1.
\end{equation}
This equation says that the restriction of $P$ along the line $\{(\kappa \cos \alpha, - \kappa \sin \alpha), \ \kappa \in \bbR \}$ ,
corresponds to the Fourier transform of $\rho_\alpha$.

Figure 5 shows how the Fourier slice theorem applies to the wavelength intensity function shown in Figure 4. Notice that different angles of integration produce different results, especially when it comes to discerning different features. Since we know beforehand that an elastic peak lies along the line $\lambda_i=\lambda_f$ we expect that an integration angle of $\alpha=\pi/4$ will produce the best results for resolving this feature as a single spike.
Since the spread of initial wavelengths is already rather concentrated, the Fourier slice theorem is often used to treat this distribution as a single point. This point, the average wavelength $\lambda_{\mrm{av}}$, is either known beforehand or deduced by projecting along the $\lambda_f$-axis. With the incident wavelength $\lambda_i=\lambda_{\mrm{av}}$ fixed, any projection can be converted to a function in terms of $\lambda_f$ only \cite{He3Apparatus}:

\begin{equation} \label{Eq:AssumeOneInRelation}
	\lambda_f=\lambda_{\mrm{av}} \cot \alpha -\lambda_{\mrm{proj}} \sec \alpha,
\end{equation}
where $\lambda_{\mrm{proj}} \in \bbR$ is a point on the line to which we project.

With this projection, we can treat the problem \eqref{Eq:PolarisationFormula2} as a one-dimensional version of \eqref{Eq:TotalSignalWavelength2piVer} with a new wavelength intensity function $\rho_{\alpha}(\lambda)$

\begin{equation} \label{Eq:TotalSignalWavelengthOneDVer}
P_\alpha(\kappa)= \int_a^b \rho_{\alpha}(\lambda) e^{2 \pi \ri \kappa \cdot \lambda} \, \D \lambda, \qquad \lambda , \kappa \in \bbR.
\end{equation}

\section{Compressed Sensing} \label{Sec:SimpleCompressed}
In this section we shall assume that we have already reduced the problem to one-dimension and write $P_\alpha, \rho_{\alpha}$ from \eqref{Eq:TotalSignalWavelengthOneDVer} as $P, \rho$.

One can discretise \eqref{Eq:TotalSignalWavelengthOneDVer} by breaking up the interval $[a,b]$ into a uniform grid of $N$ points $\lambda_{j,N}$ as in \eqref{Eq:DFTSimplifcation}, leading to the following matrix equation:
\begin{equation} \label{Eq:LinearForm}
g_j  =\sum_{k=-N}^{N}A_{j,k}f_k, \quad j=0,...,2N,  \quad g_j = \tilde{\rho}_N (\lambda_{j,N}),
\end{equation}
where $f_k = \mathrm{Constant}(k) \cdot P(k \epsilon) $ and $A$ is a DFT matrix.

Currently, to obtain the full vector $(g_j)_{j=0,...,2N}$ we need to know the entire vector $(f_k)_{k=-N,...,N}$. If we only had knowledge of a fraction of the entries of $f$ we can no longer use \eqref{Eq:LinearForm} to determine $g$ directly as the problem is now underdetermined. Therefore, the problem is not well posed and has to be modified.

The matrix equation \eqref{Eq:LinearForm} can be inverted to give
\begin{equation} \label{Eq:LinearInverse}
f_j  =\sum_{k=0}^{2N}A^{-1}_{j,k}g_k, \quad j=-N,...,N.
\end{equation}
Now suppose that $\Omega \subset \{-N,...,N\}$ denotes the of set indices corresponding to the samples of $f$ that are measured and $P_\Omega$ denotes the projection onto these samples. With this notation $P_\Omega f$ denotes the vector of samples that are measured. Therefore when we subsample from $\{-N,...,N\}$ equation \eqref{Eq:LinearInverse} becomes
\begin{equation} \label{Eq:LinearProjected}
P_\Omega f =P_\Omega A^{-1} g.
\end{equation}

The classical CS approach is to solve this problem via the now well established $l^1$ recovery problem
\begin{equation}\label{Eq:key_l1}
\min_{z \in \bbC^N} \| Wg  \|_{l^1} \quad \mbox{subject to} \quad P_{\Omega} f =P_{\Omega} A^{-1} g,
\end{equation}
where $W$ is some transformation that should make $g$ sparse. Typically this is a wavelet transformation.
We can then solve this kind of problem quickly and conveniently using convex solvers such as the (SPGL1) algorithm \cite{SPGL}.

The classical idea of CS is that $\Omega$ should be chosen uniformly at random. In this case the number of samples
$m = |\Omega|$ must satisfy
\begin{equation}\label{Eq:BasicCSGuarantee}
m \gtrsim N \cdot \mu(A^{-1} W^{-1}) \cdot s \cdot \log(N),
\end{equation}
in order to guaranty successful recovery with high probability, where $\mu(B) = \max_{1 \leq i,j \leq N}|B_{i,j}|^2$. In the case where $B = A^{-1} W^{-1}$ as above with any wavelet transform $W$ we have that $\mu(B) = 1$. In this case, as well as many others, uniform random sampling may give suboptimal results and one has to sample with (structured) variable density sampling, see \cite{LeveledCompressed} and references therein. The key problem is that the optimality of variable density sampling depends on the signal itself \cite{Roman, LeveledCompressed, WaveletSampling2}, and thus designing the best sampling pattern is a very delicate task. We will give a short demonstration below.

\subsection{How to do structured sampling right}
The key to understanding structured sampling is to understand the structure of the signal. For example, the coefficients of a signal in a wavelet basis typically have a very specific level structure. This is known as sparsity in levels.

{\bf Sparsity in levels:}
Let $x$ be a $\bbC^N$ vector. For $r \in \bbN$ let
$\mathbf{M} = (M_1,\ldots,M_r) \in \bbN^r$ with $1 \leq M_1 < \ldots < M_r$
and $\mathbf{s} = (s_1,\ldots,s_r) \in \bbN^r$, with $s_k \leq M_k - M_{k-1}$,
$k=1,\ldots,r$, where $M_0 = 0$.  We say that $x$ is
$(\mathbf{s},\mathbf{M})$-sparse if, for each $k=1,\ldots,r$,
$
\Delta_k : = \mathrm{supp}(x) \cap \{ M_{k-1}+1,\ldots,M_{k} \},
$
satisfies $| \Delta_k | \leq s_k$.
This known structure can be utilised when designing the sampling strategy and is the motivation behind multilevel sampling.

{\bf Multilevel sampling:}
\label{multi_level_dfn}
Let $r \in \bbN$, $\mathbf{N} = (N_1,\ldots,N_r) \in \bbN^r$ with $1 \leq N_1
< \ldots < N_r$, $\mathbf{m} = (m_1,\ldots,m_r) \in \bbN^r$, with $m_k \leq
N_k-N_{k-1}$, $k=1,\ldots,r$, and suppose that
$
\Omega_k \subseteq \{ N_{k-1}+1,\ldots,N_{k} \},\quad | \Omega_k | = m_k,\quad
k=1,\ldots,r,
$
are chosen uniformly at random, where $N_0 = 0$.  We refer to the set
$
\Omega = \Omega_{\mathbf{N},\mathbf{m}} = \Omega_1 \cup \ldots \cup \Omega_r
$
as an $(\mathbf{N},\mathbf{m})$-multilevel sampling scheme.
The key is that in the case of Fourier sampling, represented by the $B$ above, combined with a wavelet transform $W$ such that the recovery problem becomes \eqref{Eq:key_l1}, the multilevel sampling should match the level structure of the wavelets. More precisely,  $\mathbf{N} = \mathbf{M}$. In this case, if $x$ is $(\mathbf{s},\mathbf{M})$-sparse with total sparsity $s = s_1 + \hdots + s_r$ and $s_1 = M_1 = m_1$. Then the total number of samples needed is
\begin{equation} \label{Eq:MultilevelSamplingRule}
m = m_1 + \hdots +m_r \gtrsim s_1 + (s_2 + s_3 + \hdots + s_r) \log(N).
\end{equation}
In particular, by utilizing the level structure in the sampling, one can outperform the standard CS results. For a more in-depth analysis and explanation see \cite{Foundations, LeveledCompressed, WaveletSampling2}. See also \cite{Donoho,Candes_PNAS} for early versions of this kind of sampling.

\section{Continuous Compressed Sensing} \label{Sec:Compressed4Continuous}
The motivations behind continuous CS are: (i) to obtain a continuous approximation in the CS reconstruction, as opposed to a discrete approximation, as this allows for an easy non-linear change of variables  to obtain the scattering function and the intermediate scattering function. (ii) If an alternative basis to the Fourier representation yields a better representation of the function to be recover, one wants the freedom to use that. In particular, one can try approximating the wavelength intensity function $\rho$ in terms of a new \emph{Reconstruction Basis} $\sigma_n, \ n \in \bbN$:
\begin{equation} \label{Eq:FunctionApproximation2}
\tilde{\rho}(\lambda) \approx   \sum_{n =1}^N \langle \rho , \sigma_n \rangle \sigma_n(\lambda), \quad \lambda \in [a,b].
\end{equation}
For technical reasons, one often requires these functions to form an orthonormal basis of $L^2[a,b]$, e.g. Legendre polynomials, splines, wavelets etc., although this condition can be relaxed to other groups of functions like frames \cite{Mallat}. For this paper we shall be using Daubechies wavelets \cite{DaubechiesCPAM} exclusively as our reconstruction basis. Let us quickly discuss \emph{why} one would want to work with another basis.

Apart from the benefit of keeping the problem continuous, one also has the freedom to \emph{choose} which basis $\sigma_n$ to work with, making the approach more versatile than a straight DFT approach, where we are essentially forced to work with a pixel basis every time.

Furthermore, the notion of \emph{sparsity} is now in terms of the coefficients $\langle \rho , \sigma_n \rangle$, which means we have the additional advantage of choosing a basis that makes the function $\rho$ sparse. As we shall see, this opens up the possibility of using compressed sensing where traditional sparsity does not hold. In addition, this approach is closer to the philosophy that $\rho$ being sparse should relate to $\rho$ having low information content; a choice of basis $\sigma_n$ that makes $\rho$ sparse tells us how to (approximately) express the function $\rho$ with a few non-zero coefficients.

Since one is still sampling data that corresponds to Fourier coefficients of $\rho$, it is impossible to exclusively work with their choice of basis $\sigma_n$. Instead one has to convert Fourier series coefficients into coefficients in the basis $\sigma_n$. This is achieved by working with the infinite change of basis matrix for the two bases:

\begin{equation} \label{Eq:ContinuousChangeOfBasis2}
B_{k,n}= \langle \sigma_n, \chi_{k, \epsilon} \rangle, \quad n \in \bbN, \ k \in \bbZ.
\end{equation}
As opposed to \eqref{Eq:key_l1}, we now end up solving the infinite dimensional convex optimisation problem of finding
\begin{equation} \label{Eq:ConstrainedLinearEll1Continuous}
\begin{aligned}
h^* \in \operatorname*{argmin}_{h \in \ell^2(\bbN)} \Big\{ \|h\|_1 : P_\Omega B h = P_\Omega f \Big\},
\end{aligned}
\end{equation}

Again $P_\Omega$ denotes the projection onto the samples we have taken. In practice we cannot solve for the infinite solution to \eqref{Eq:ConstrainedLinearEll1Continuous}, therefore we truncate the reconstruction basis in a similar fashion to how we truncate the Fourier basis. This means we end up computing
\begin{equation} \label{Eq:ConstrainedLinearEll1ContinuousTruncated}
h^* \in \operatorname*{argmin}_{h \in \bbC^N} \Big\{ \|h\|_1 : P_\Omega B P_N h = P_\Omega f \Big\},
\end{equation}
where $P_N$ denotes the projection onto the first $N$ functions in the reconstruction basis. This problem is now numerically feasible since the submatrix $P_\Omega B P_N$ is now finite (see \cite{LeveledCompressed} for estimates on how to choose $N$). The solution to \eqref{Eq:ConstrainedLinearEll1ContinuousTruncated}, let's say $h^*$, is recognised as the (approximate) wavelet coefficients of the intensity function. We can then use these wavelet coefficients to compute an approximation to $\rho$ evaluated at any point on the interval $[a,b]$ by following \eqref{Eq:FunctionApproximation2}:

\begin{equation} \label{Eq:EvaluateFunctionAfterSolving}
\rho(\lambda) \approx   \sum_{n =1}^N h^*_n \sigma_n(\lambda), \quad \lambda \in [a,b].
\end{equation}

From here one can use the same multilevel sampling techniques as the discrete case to reconstruct $(\mb{s},\mb{M})$-sparse coefficients. Moreover, the sampling rule \eqref{Eq:MultilevelSamplingRule} also applies in this case.

Note that this approach is very different from \eqref{Eq:key_l1}. First, \eqref{Eq:ConstrainedLinearEll1ContinuousTruncated} is not based on the traditional DFT, however, fast $n \log(n)$ implementation of matrix vector multiplication with $P_\Omega B P_N$, the finite section of the infinite matrix $B$, is possible. Second, the  solution to \eqref{Eq:key_l1} would (modulo the wavelet "crime", see \cite{ContinuousCompressed}) yield wavelet coefficients of a discretisation of the truncated Fourier series and are therefore not wavelet coefficients of the original function $\rho$. For further explanation and numerical examples demonstrating the benefits the continuous approach and the differences with the discrete approach, see \cite{ContinuousCompressed,LeveledCompressed,AHBogdanTeshckeGenSamp}.

\section{CS for Phonon Detection} \label{Sec:AppliedCSPart1}

\begin{figure}[t]
\centering
\includegraphics[width=0.92\textwidth]{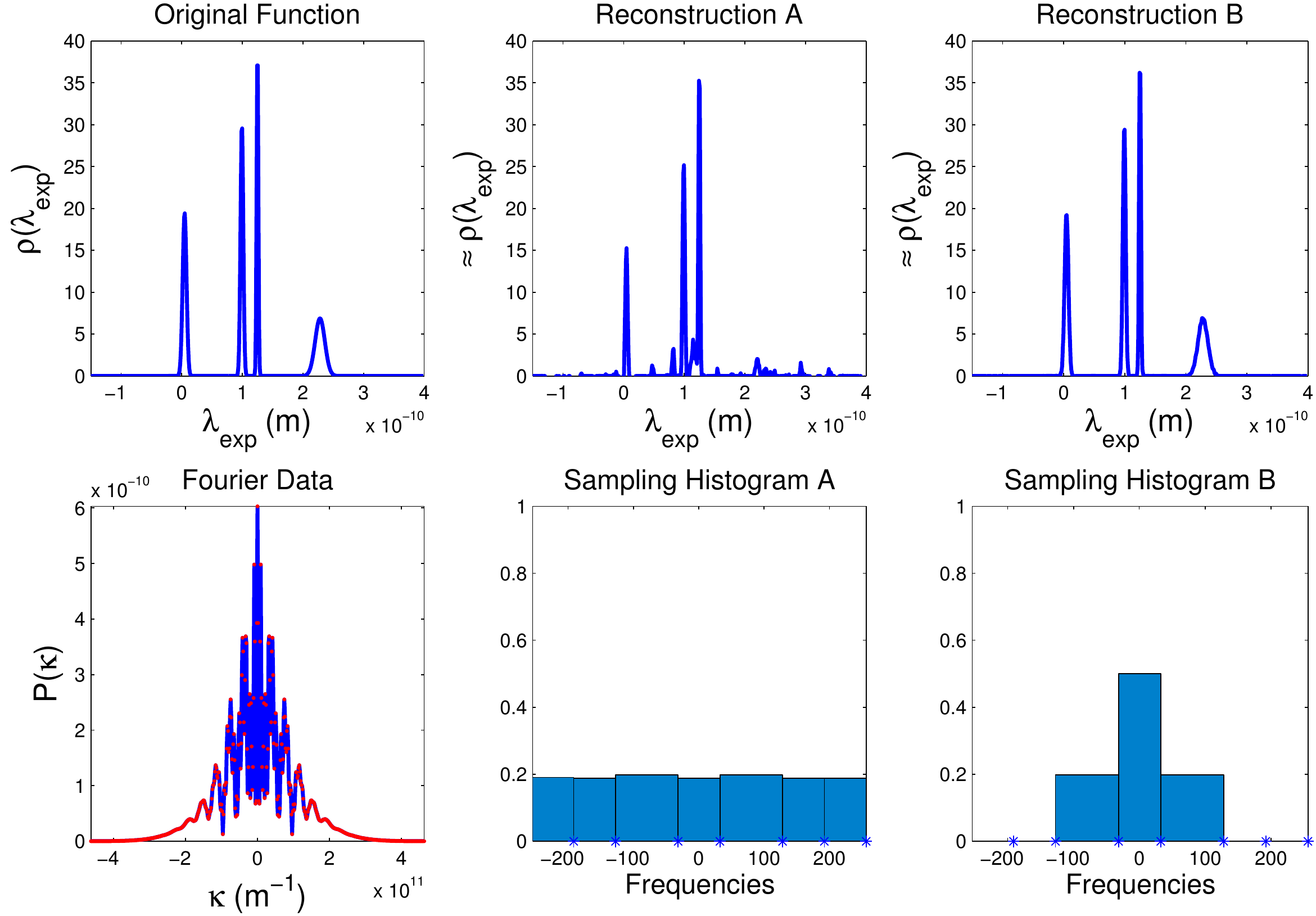}

\vspace{10pt}

\caption{Demonstrating CS for the 1D $45$ degree projection shown in Figure 5, using uniform and multilevel sampling. Samples are taken from the Fourier data according to the sampling histograms shown. Sampling pattern A is unreliable in reconstructing the rightmost feature as it is the least sparse of the four peaks while sampling pattern B remedies this by taking more of the lower frequency values that it depends upon. Reconstructions are at a resolution of 512 data points.}
\label{Fig:2DSimCS}
\end{figure}

In this section we look at the performance of the CS approach described in the previous section by looking at examples of phonon detection. We shall first look at its effects on the one-dimensional projections shown previously and then focus on a real $^3$He spectrum for scattering of gold where more exotic signal behaviour is present.

\subsection{Simulated 1D Example} \label{SubSec:Simulated1D}

For consistency with previous sections we shall first work with the 45 degree projection shown earlier. Although this is a simplified model it clearly demonstrates some of the basic properties of compressed sensing. Reconstructions are shown in Figure 6.

Recall that we project along a 45 degree angle in an attempt to reduce the spread caused by the inaccuracy of the wavelength of the incident He$^3$ beam. In particular, since we know that there will always be an elastic feature in the wavelength intensity function which itself is slanted at a 45 angle, this choice is seen as ideal for refocusing the various phonon features to be closer to that of a delta spike. Not only is this useful in preventing features from overlapping each other but this also increases sparsity which is ideal for compressed sensing; if features are sparse then by the rule \eqref{Eq:BasicCSGuarantee} we can subsample to a great degree since the signal itself is very sparse. In this case, a change of basis may not be needed.

If one goes for a uniformly random approach to subsampling (as in Reconstruction A), as opposed to multilevel sampling discussed above, then there is only so far that one can go  before problems occur. At around 20\% subsampling, the reconstruction becomes unreliable in recovering the least sparse of the features on the far right. Around 30\% however, the rightmost features is typically reconstructed but it is nonetheless unreliable. There is however an even more effective way of reliably reconstructing the rightmost feature by using multilevel sampling (as in Reconstruction B).
The theory on how to design optimal structured multilevel sampling strategies is very new \cite{Roman, LeveledCompressed, WaveletSampling2} and this is a highly unexplored topic. We do not attempt to seek optimality here, as this paper is about establishing the effectiveness of CS in HeSE. Note that, since there is no wavelet change of basis ($W = I$, the identity, in \eqref{Eq:key_l1}) in this case, the theoretical understanding of the effect of multilevel sampling is not fully understood. This is under investigation together with optimality conditions.

\subsection{Real Phonon Spectrum}  \label{SubSec:RealGold1D}

As we have already mentioned, real phonon spectra contain more exotic features than the simulation given in the previous example. Naturally noise adds to the data due to some experimental uncertainties in the apparatus, but more unusual are the relative sizes and shapes of the various features.

\begin{figure}[t]
\centering
\includegraphics[width=0.92\textwidth]{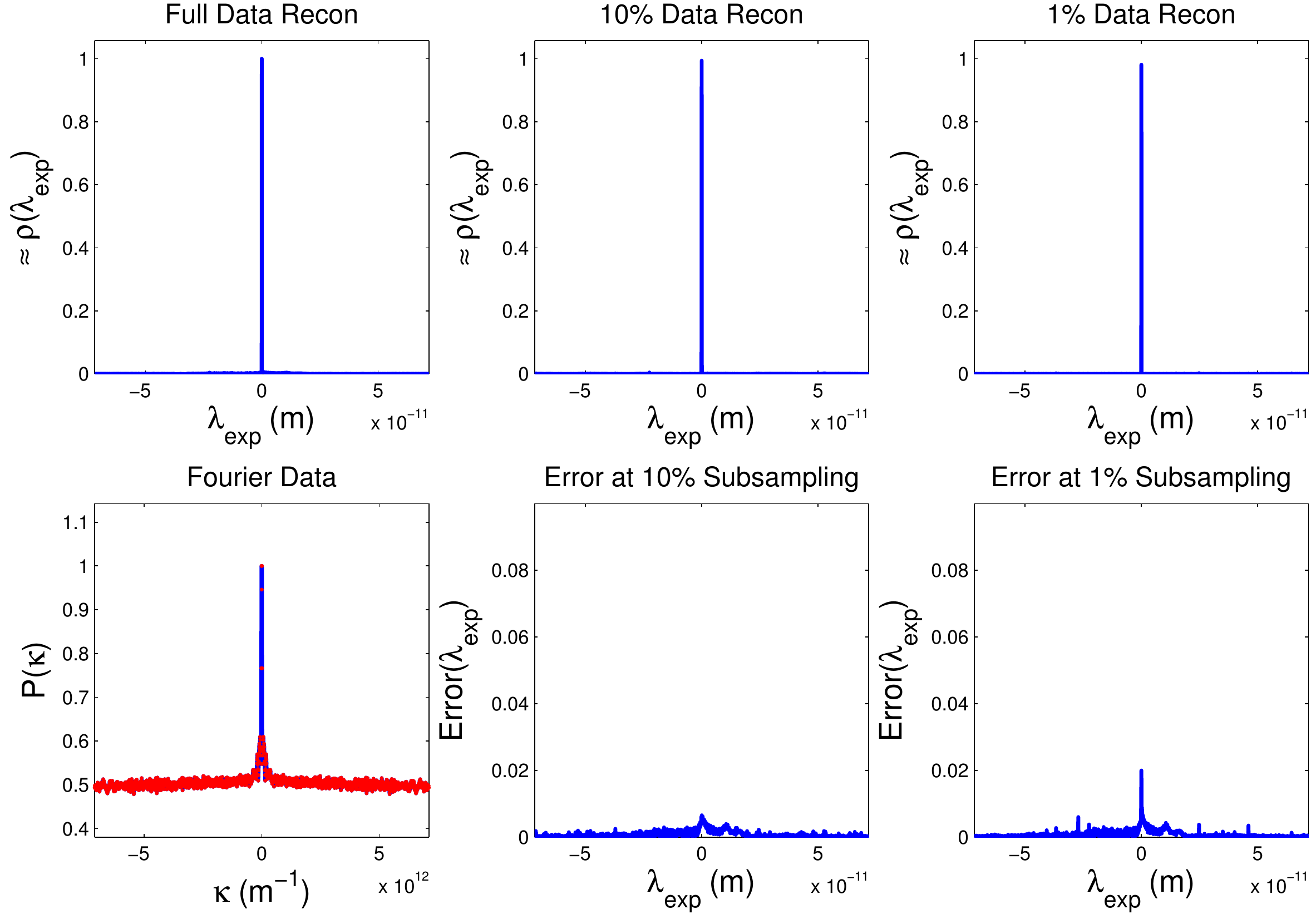}

\vspace{10pt}

\caption{Compressed sensing reconstructions for a gold phonon spectrum (a.u.). With this choice of viewing range, only the elastic peak can be clearly seen. For a zoomed in closeup on other features see Figure 8. In this Figure the sampling is performed uniformly at random which recovers the highly sparse central peak. This is suboptimal compared to the multilevel sampling in Figure 8. Frequencies sampled are from the range $\{-1024,...,1023\}$ and reconstructions are at a resolution of 2048.}
\label{Fig:GoldCS}
\end{figure}

In Figure 7 we have uniform sampling reconstructions for a typical gold spectrum (for more details see Supporting Information) with projection at 45 degrees to focus on the elastic peak, which is the only clearly visible feature in the graphs. The peak is extremely fine and is in fact even smaller than the pixel resolution used for reconstruction (2048) which can be determined from the observation that the Fourier data has yet to decay to zero near the highest frequencies. Consequently this is an ideal situation for CS since this feature is almost as sparse as can be. Hence, one can subsample to a much greater degree (e.g. 1\%) than in the previous example.

However, what has happened to the other non-elastic features in this spectrum? At first one might come to the conclusion that they are not there at all, however, focusing on a small part of this spectrum reveals features that are over 200 times smaller than the measured intensity of the large elastic spike. Figure 8 shows various CS reconstructions zoomed in on this region. Notice that we still have the elastic peak visible, along with a couple of more (and less sparse) features. Like in the previous example, we expect the smooth features to be more dependent upon the lower frequency samples. Therefore, when we attempt to take just the first 10\% of samples all from the lowest frequencies (the \emph{linear} reconstruction) we find that these features are at the very least present, unlike the 10\% uniform sampling approach where only the central peak remains. On the other hand, the central peak suffers from Gibbs artifacts which manifests themselves as wave like features near the central peak as well as broadening of the peak itself.

\begin{figure}[t]
\centering
\includegraphics[width=0.92\textwidth]{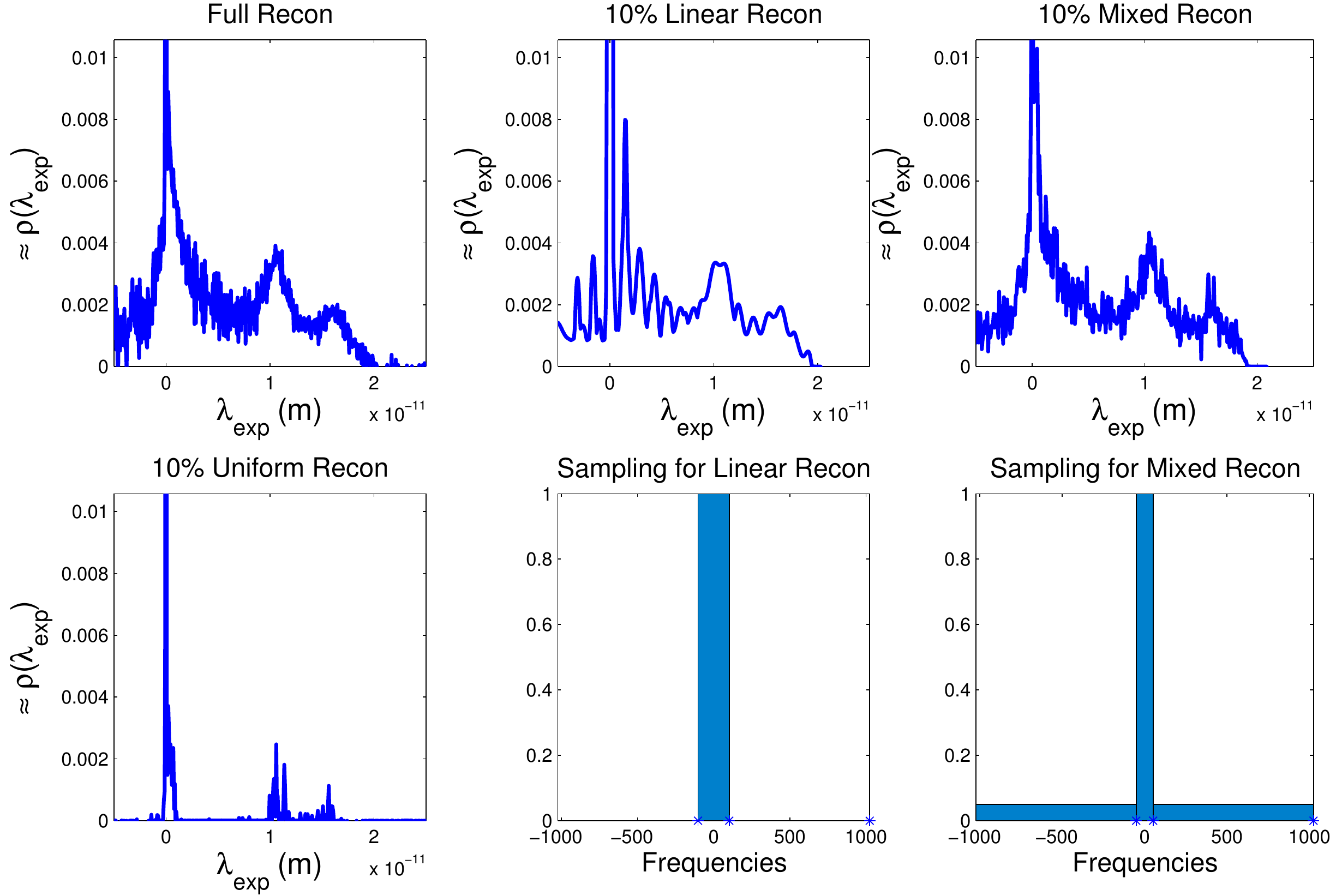}

\vspace{10pt}

\caption{Compressed sensing reconstructions for the same gold phonon spectrum, zoomed in so that features beside the elastic peak are visible. For a complete picture of the spectrum consult Figure 7. Notice that the smaller features shown here are over 200 times smaller than the elastic peak. Reconstructions shown here are not only uniform (as in the bottom left graph) but also linear (i.e. straight Fourier series) and non-linear examples using roughly the same number of samples across each.}
\label{Fig:GoldZoomedCS}
\end{figure}

Instead, one can opt for a mix of these methods by taking the first 5\% of samples from the lowest frequencies and the other 5\% taken uniformly from the rest. This approach empirically performs the best out of the three methods, resolving the low resolution features without the Gibbs artefacts of the linear approach. Note that, as demonstrated in \cite{LeveledCompressed, Roman}, the optimal sampling procedure is signal structure dependent. How to choose optimal multilevel sampling is beyond the scope of this paper, and we have deliberately chosen a simple two-level sampling pattern, which is a reasonable all rounder, in order to demonstrate the effectiveness of the sampling technique.

\subsection{Comparing CS Techniques} \label{Sec:AppliedCSPart2}

In this section we look at an example of how the continuous wavelet approach to compressed sensing can be used to tackle problems that are beyond the capabilities of the traditional compressed sensing approach described earlier.

\begin{figure}[t]
\centering
\includegraphics[width=0.92\textwidth]{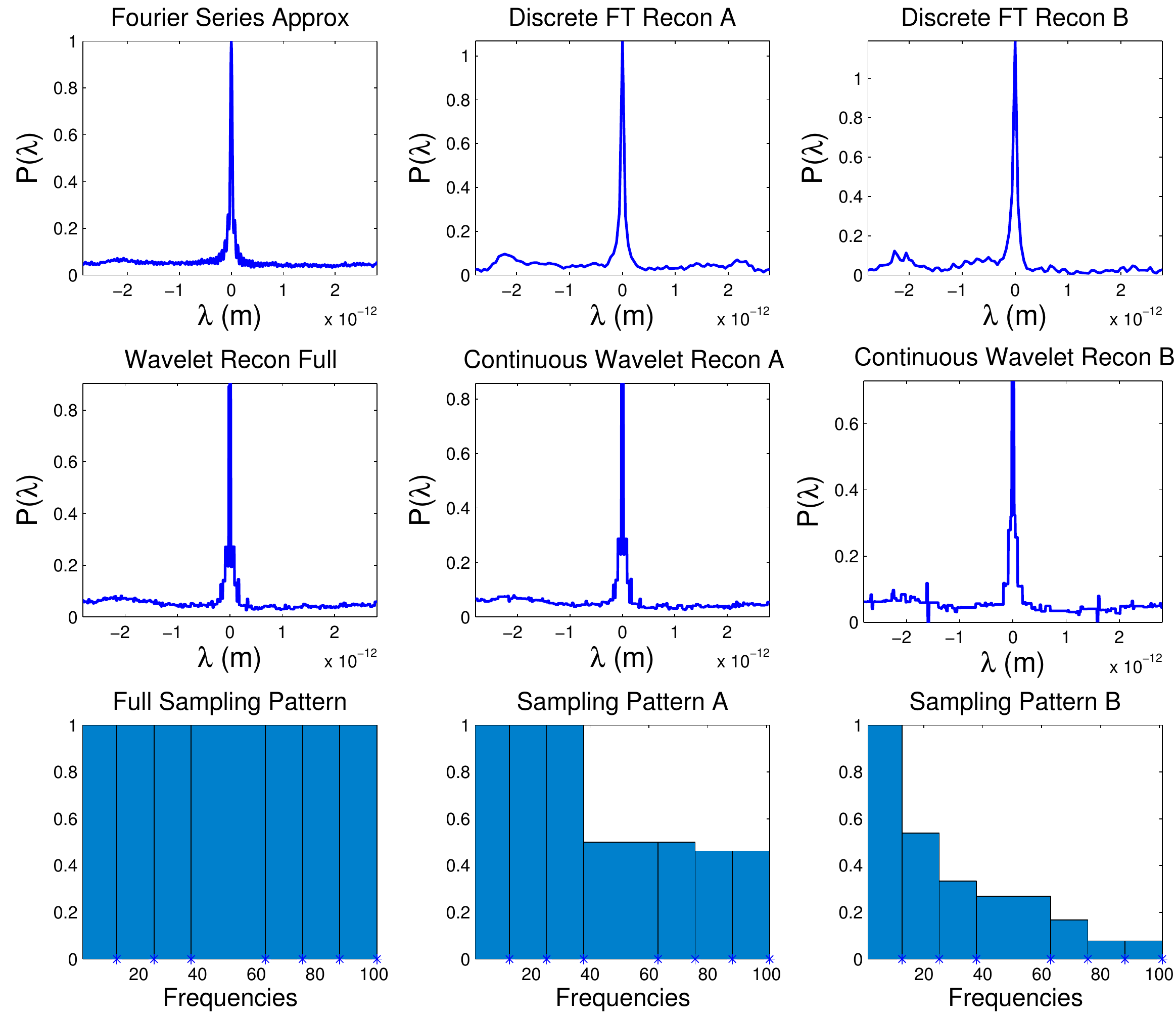}

\vspace{10pt}

\caption{(Continuous vs discrete CS) Reconstructions of a spin-echo spectrum (a.u.) of CoPc molecules deposited on Ag(001) (see Supporting Information for details on the sample preparation) with a noticeable baseline feature. In the DFT reconstructions the baseline level of around 0.1 is no longer flat leaving bumpy artefacts while the wavelet reconstructions preserve this flat feature. As we are only subsampling from 101 frequency points, considerable Gibbs artefacts are present in the Fourier series approximation. The DFT reconstructions have a resolution of 101 points, while the continuous Fourier series and wavelet reconstructions have been rasterised at a resolution ten times this number.}
\label{Fig:WaveletAdvantage}
\end{figure}

Prior to data acquisition the spacing in current (equivalent to spacing in $\kappa$) must be chosen, which in turn determines the length of the wavelength window $[a,b]$ that the wavelength intensity function $\rho$ is constructed over. If $\rho$ is not truly supported on this window, then by \eqref{Eq:PeriodicWavelength} we instead reconstruct the periodised version of $\rho$.  In particular, if peaks in the intensity function decay particularly slowly relevant to the window then the intensity function will stay considerably above zero throughout that window. Because of this, the traditional compressed sensing approach applied earlier cannot be used successfully here as the function is maximally non-sparse.

However, if one recalls the wavelet reconstruction bases that are used, then one quickly notices that they both have a constant function as the first basis function. This effectively means that the base level caused by slow decay is captured by this single basis function, which keeps the function sparse in these bases.

Figure 9 compares the two compression techniques for the diffusion of cobalt phthalocyanine (CoPc) on Ag(001) with an observable baseline feature. The full set of polarisation data points only corresponds to the first 101 frequencies and therefore there is noticeable Gibbs artefacts around the elastic peak in the Fourier series approximation. This strongly suggests that the Fourier series approximation here is not a particularly accurate approximation to the true intensity function.

Furthermore, the wavelet approximation aims to reconstruct the true underlying continuous wavelength intensity function, unlike the DFT approach which attempts to reconstruct a discretised form of the Fourier series approximation. Consequently,  even with full sampling, the wavelet reconstruction is noticeably different to  the Fourier series approximation. This reflects the fact that, as we are handling real data, we cannot directly compare to the true underlying wavelength intensity function.

Regardless we clearly observe that the baseline feature is preserved under subsampling using wavelets where the DFT approach clearly fails, matching predictions based on sparsity observations earlier. Note that both techniques use exactly the same samples.
While we are able to subsample to a reasonable degree here ($\approx 33\%$), one should ideally work with a larger range of frequencies to truly exploit the benefits of this approach, i.e. subsampling from Polarisation data with thousands of points rather than hundreds.

\section{Molecular Diffusion Using Continuous CS} \label{Sec:BackGrdHASPrt2}

In this section we shall focus on studying diffusive properties of surfaces. As we mentioned earlier, if one wants to determine how effective a surface behaves, e.g. for catalysis, then one must study how molecules move on top of the surface over time. This is very different to the previous phonon examples covered earlier because we are no longer just considering the motions of nuclei in the lattice. Instead we have species (molecules, atoms) diffusing on top of the surface that are interacting with each other.

Three models of diffusion are given in Figure 10. An important question is how one can differentiate these three types of diffusive regimes on a surface by using He atom scattering.

\subsection{Scattering Cross Sections and the Van Hove Formalism} \label{SubSec:VanHove}

The van Hove formalism was initially developed for thermal neutron scattering \cite{Squires} and we shall mention a few key results from this theory. This theory establishes a relationship between the neutron spectra (as a function of energy and momentum transfer) and the dynamics of the nuclei within the sample. This theoretical approach makes some assumptions that generally holds well for neutrons, for example
\begin{itemize}
\item The incoming beam of neutrons has a fixed incident wavevector $\mb{k}_i$ (and therefore a fixed incident energy). This is equivalent to monochromatisation of the beam.
\item The potential for scattering of each nuclei is modelled as a Fermi pseudopotential, which is a delta spike at its position. This does \emph{not} hold well for $^3$He.
\item A scattered neutron interacts with the bulk potential at most once. This is carried through to $^3$He scattering (considering multiple scattering events, while possible,  is very challenging and often avoided \cite{QuasiNeutron}).
\end{itemize}
With these assumptions one can compute the differential scattering cross section $\frac{\D^2 \sigma}{\D \Omega \D (\hbar\omega)}$ , defined as the number of neutrons scattered through a solid angle $\D \Omega$ with change of energy $\D (\hbar\omega)$ divided by the flux of incident neutrons \cite[Eq.  4.13]{Squires}:

\begin{figure}[t]
\centering
\includegraphics[width=0.92\textwidth]{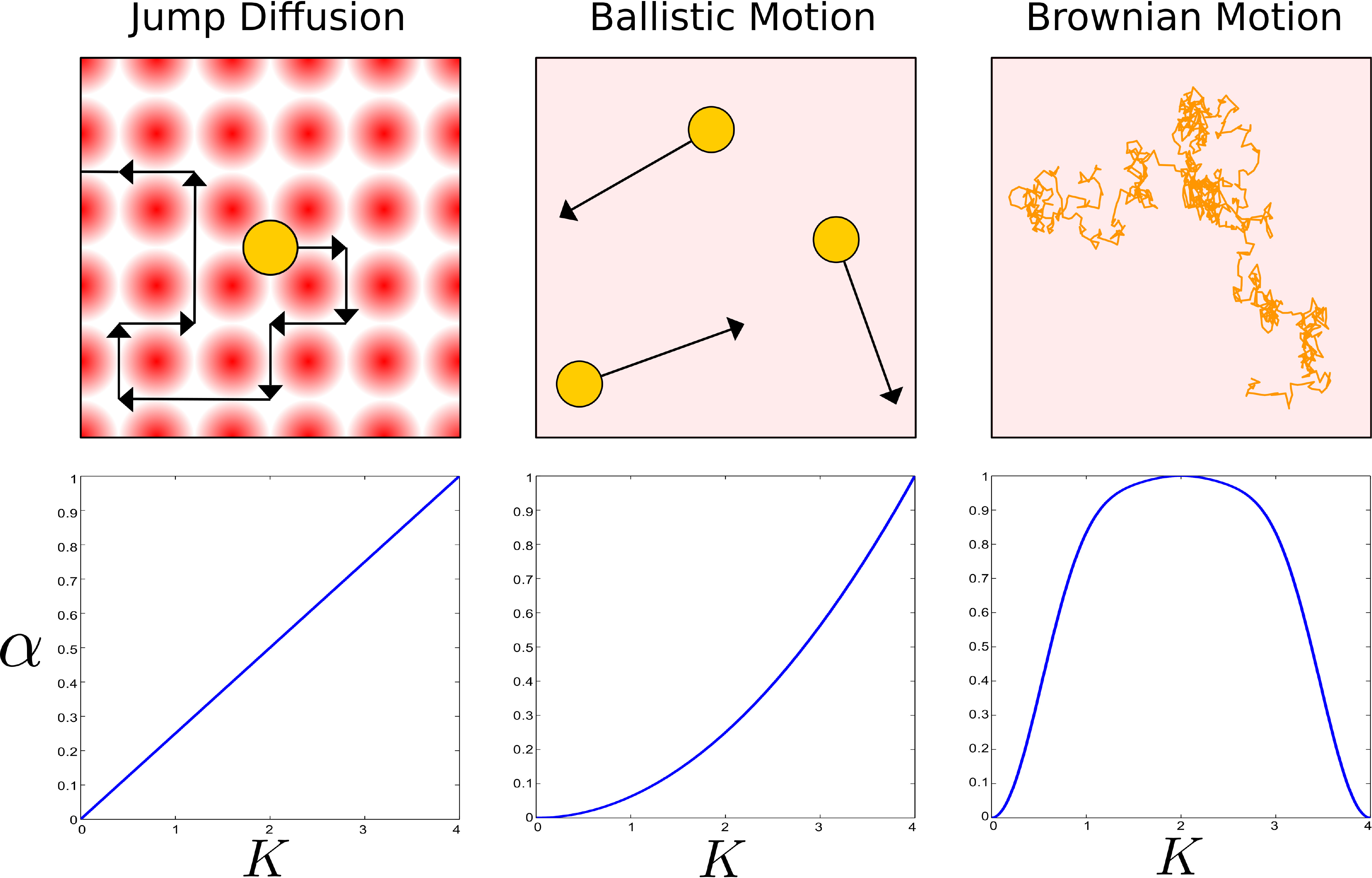}

\vspace{10pt}

\caption{A depiction of three diffusive regimes. Orange denotes molecules on the surface and red denotes the surface potential. In the jump diffusion case the molecules are assumed to move instantly between vacant sites where the potential energy is smallest. The corresponding dephasing rates are shown in the lower plots.}
\label{Fig:DiffusionTypes}
\end{figure}
\begin{equation} \label{Eq:ScatteringCrossSectionFormula}
	\left( \frac{\D^2 \sigma}{\D \Omega d(\hbar\omega)} \right)= \frac{\sigma}{4\pi} \frac{k_f}{k_i} N S(\mb{Q},\omega),
\end{equation}
where $\sigma$ is the total scattering cross section, $k_i=|\mb{k}_i|, k_f=|\mb{k}_f|$ are the magnitudes of the incoming and outgoing scattering wavevectors, $\mb{Q}=\mb{k}_i-\mb{k}_f$, $N$ is the total number of nuclei, $\hbar\omega$ denotes the change in energy (from \eqref{Eq:FrequencyEnergyRelation}) and $S(\mb{Q},\omega)$ is called the \emph{Scattering Function} (SF). What makes \eqref{Eq:ScatteringCrossSectionFormula} particularly useful is that the structure factor can be related to the motion of nuclei in the bulk via multiple Fourier transforms:

\begin{equation}\label{Eq:Correlation2StructureFactor}
\begin{aligned}
G(\mb{R},t) &= \int_{\bbR^3} \exp( -2 \pi \ri  \mb{Q} \cdot \mb{R}) I(\mb{Q},t) \, \D \mb{Q}
\\ & = \int_{\bbR^3} \int_{\bbR}  \exp( -2 \pi \ri ( \mb{Q} \cdot \mb{R}+ \omega t)) S(\mb{Q}, \omega) \, \D t \, \D \mb{Q}.
\end{aligned}
\end{equation}
The function $G(\mb{R},t)$ is often called the \emph{van Hove correlation function} and $I(\mb{Q},t)$ the \emph{Intermediate Scattering Function} (ISF). The correlation function $G(\mb{R},t)$ can be interpreted as the probability that a particle will be at position $\mb{R}$ at time $t$ provided this particle or a different one is at the origin at time $t=0$. $G(\mb{R},t)$  contains information about both self- and collective diffusion. However, if the spatial correlations between particles are negligible, the scattering function is essentially sensitive to the dynamics of single particles (self-diffusion).

\begin{figure}[t]
\centering
\includegraphics[width=0.85\textwidth]{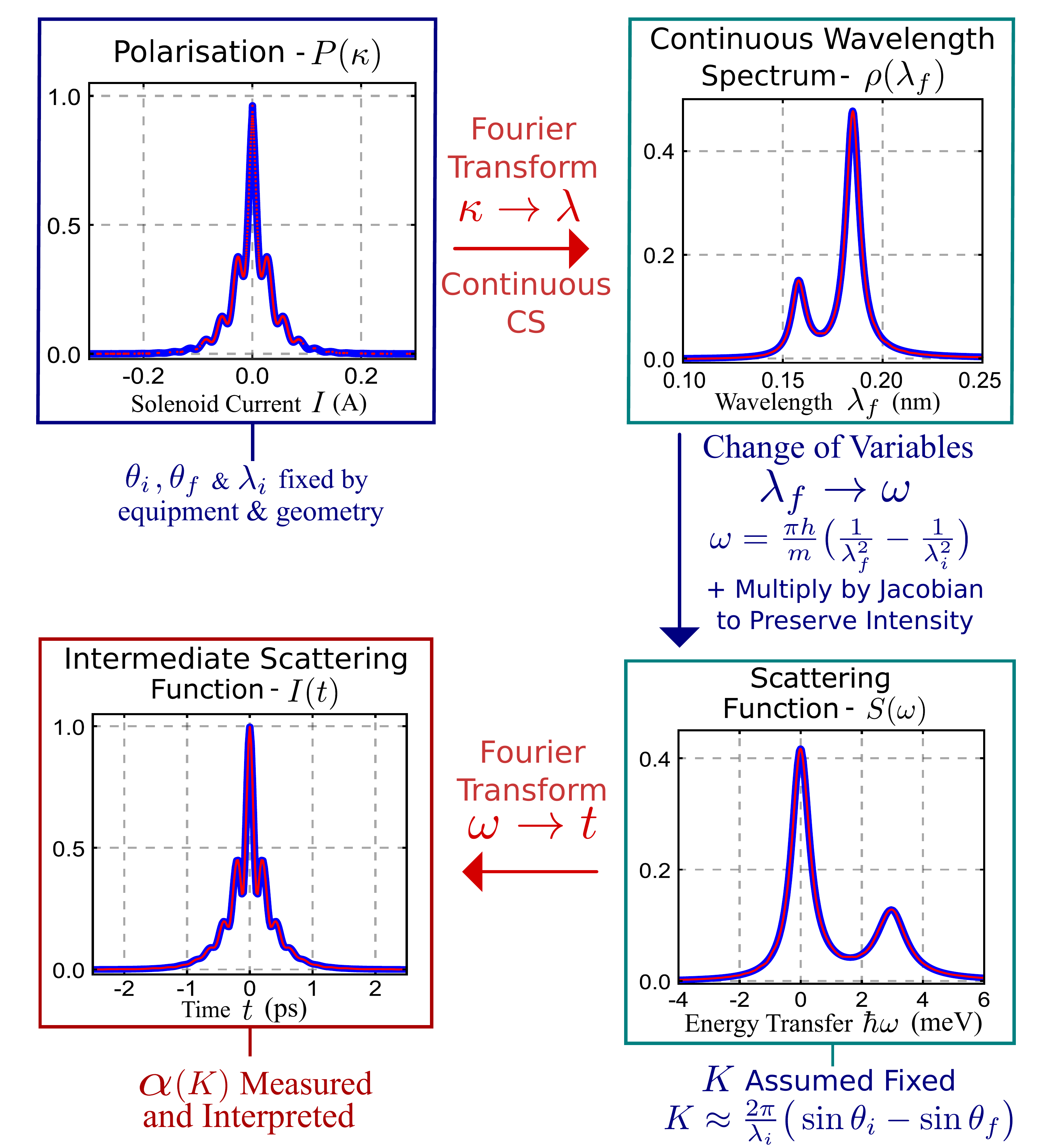}

\vspace{10pt}

\caption{A detailed 1D description of the full loop presented in Figure 1. Blue lines denote the true underlying signal and red lines denote the reconstructions (except for the top left where the red dots denote the sampling points). Sampling points are taken according to the sampling histogram present in Figure 1. Note that there is no comparison with standard discrete CS here as the change of variable technique is impossible in the discrete setup. This can only be done in the continuous case.}
\label{Fig:1DLoop}
\end{figure}

\subsection{Extension to HeSE} \label{SubSec:Dephasing}
Assuming that there there is no correlation between particle dynamics, formula \eqref{Eq:ScatteringCrossSectionFormula} can be adapted \cite{He3Apparatus} to the surface sensitive helium scattering approach using

\begin{equation} \label{Eq:ScatteringCrossSectionFormulaHAS}
	\frac{\D^2 \sigma}{\D \Omega \D \omega} =  S(\mb{K},\omega) \cdot |F(\mb{K},\omega)|^2,
\end{equation}
where the change of momentum $\mb{K}$ (the $\Delta$ is omitted) now denotes $\mb{k}_f-\mb{k}_i$ projected onto the surface (like in \eqref{Eq:Conservation}) and $F(\mb{K}, \omega)$ denotes the \emph{Form Factor} that takes into account the fact that $^3$He atoms are scattered from the electronic cloud of the atoms/molecules. This contrasts with the nuclear scattering of neutrons. For simplicity, we shall be assuming that the form factor can be compensated for when interpreting the scattering data, which effectively means setting this term to be equal to 1. Similarly \eqref{Eq:Correlation2StructureFactor} still holds but for the two-dimensional equivalents of $G(\mb{R},t), I(\mb{Q},t)$ and $S(\mb{Q},t)$ (with $\mb{Q} \in \bbR^3$ replaced by $\mb{K} \in \bbR^2$  and now $\mb{R} \in \bbR^2$)

As we shall see shortly in the next section, one often does not have the luxury of knowing the SF $S(\mb{K}, \omega)$ on the entire $(\mb{K}, \omega)$ space, likewise for the ISF $I(\mb{K}, t)$. This means that one cannot easily calculate the van Hove correlation function directly. Instead, it is sometimes preferable to infer properties of $G(\mb{R},t)$ from partial knowledge of the SF/ISF. One such example is the \emph{Dephasing Rate} $\alpha(\mb{K})$, which describes the decay in $t$ of the ISF as a function of $\mb{K}$. Formally,
\begin{equation} \label{Eq:DephasingRateDef}
\alpha(\mb{K})= \inf \{t>0 : |I(\mb{K},t)|=e^{-1}|I(\mb{K},0)| \}.
\end{equation}
Different types of models for self-diffusion are presented in Figure 10. For the Brownian/Ballistic/Jump Motion the dephasing rate shows a quadratic/linear/periodic dependence upon $\mb{K}$ \cite{He3Apparatus}.

\subsection{Changing Variables} \label{SubSec:Wavelength2Energy}

Now we connect up the dots between the van Hove formalism described in the last section and the experimental background.

Recall that by \eqref{Eq:ScatteringCrossSectionFormulaHAS} we know that the scattering cross section is expressed in terms of change of surface momentum $\mb{ K}$ and energy $ E=\hbar \omega$. Therefore if one wants to convert wavelength intensity to intensity in terms of energy or wavelength one first needs to change variables (recalling \eqref{Eq:Conservation} and \eqref{Eq:KineticChange}):
\begin{equation}
\begin{aligned} \label{Eq:NonlinearChangeOfCoordinates}
\omega & = \frac{E}{\hbar}  = \frac{\hbar}{2m} k_f^2-\frac{\hbar}{2m} k_i^2=\frac{ \pi h}{m} \left( \frac{1}{\lambda_f^2}-\frac{1}{\lambda_i^2} \right)
\\  K & =k_f \sin \theta_i - k_i \sin \theta_f = \frac{2 \pi}{\lambda_f} \sin \theta_i - \frac{2 \pi}{\lambda_i} \sin(\gamma - \theta_i).
\end{aligned}
\end{equation}
Here $\gamma=\theta_i+\theta_f$ is the total scattering angle between the source/surface/detector setup which cannot be changed for the Cambridge spin-echo apparatus. Instead one tilts the surface in order to vary the incident angle $\theta_i$, which in turn determines $\theta_f$. The direction of the change in surface momentum $\mb{K}$ is determined by the geometry of the apparatus but is always parallel to the surface (formally when we restrict $\mb{K}$ to the plane in the source/surface/detector setup it becomes a scalar hence why we only have a scalar $ K$ in \eqref{Eq:NonlinearChangeOfCoordinates}).

From \eqref{Eq:NonlinearChangeOfCoordinates} we see that the initial scattering angle $\theta_i$ is also an important variable in our experiments, therefore it is convenient to explicitly declare this dependency by rewriting $\rho(\lambda_i,\lambda_f)$ as $\rho(\lambda_i,\lambda_f,\theta_i)$. Our goal is to convert knowledge of $\rho(\lambda_i,\lambda_f,\theta_i)$ to knowledge of the scattering cross section/SF and then to the ISF. In this paper we focus on the approach where we fix $\lambda_i$ (using a Fourier slice) and $\theta_i$, leaving a function of one variable $\rho(\lambda_f)$, like in the phonon case.

By fixing $\lambda_i, \theta_i$ we only know $\tilde{S}(K,\omega)$ on a one-dimensional path in $(K,\omega)$ space. Because of this issue, some choose to exclusively work with the energy $\omega$. With this approach the wavelength intensity function is converted into a frequency intensity function, which we interpret as $S(\omega)$. This involves using the change of variables \eqref{Eq:NonlinearChangeOfCoordinates} along with a Jacobian term to preserve intensity:
\begin{equation} \label{Eq:NonLinearWithJacobian4LambdafOnly}
	S(\omega(\lambda_f))   =\rho(\lambda_f) \cdot \left( \frac{\D \omega}{\D \lambda_f}(\lambda_f) \right)^{-1}.
\end{equation}
From here one can Fourier transform $S(\omega)$ to derive an approximation to the ISF $I$ in \eqref{Eq:Correlation2StructureFactor} as in  \cite{He3Apparatus}. One problem however, is that the paths taken in $(K,\omega)$ space with $\theta_i,\lambda_f$ fixed are not straight lines parallel to the $K$-axis which otherwise would have justified this method using the Fourier slice theorem. Because of this issue we refer to $I(t)$ as an \emph{approximate-ISF}.

\section{The Big Picture} \label{Sec:FullLoop}

Let us sum up the various stages we have discussed. The three key steps in order are:

\begin{enumerate}
\item The initial Fourier transform measurement step \eqref{Eq:TotalSignalWavelengthOneDVer} from polarisation to wavelength that is subject to continuous CS.
\item A non-linear change of variables \eqref{Eq:NonlinearChangeOfCoordinates} from wavelength to energy.
\item A further Fourier transform step \eqref{Eq:ScatteringCrossSectionFormulaHAS} from energy to time, generating an approximation to the Intermediate Scattering Function.
\end{enumerate}

Figure 11 shows these steps in more detail. After the Intermediate Scattering Function is reconstructed the dephasing rate can be extracted and the diffusive properties analysed. There is no comparison with standard discrete CS here as the change of variable technique is impossible in the discrete setup.
As discussed in the introduction, the change of variables \eqref{Eq:NonlinearChangeOfCoordinates} is incompatible with the standard DFT based CS methods since the uniform grid of coordinates in wavelength space is transformed into a non-uniform grid in $(\mb{K},  E)$-space. With continuous CS one can work backwards, first specifying a uniform grid in $(\mb{K},  E)$-space which is converted to a non-uniform grid in wavelength space. Since,  using the continous CS approach, the reconstructed solution of (\ref{Eq:ConstrainedLinearEll1Continuous}) is a function rather than a vector, one can directly sample the wavelength intensity function on this non-uniform grid directly and use   \eqref{Eq:NonLinearWithJacobian4LambdafOnly} to compute the scattering function.

\section{Conclusion and Outlook} \label{Sec:Conclusion}

This work demonstrates that continuous compressed sensing can be used to reduce measurement times by at least an order of magnitude whilst capturing both phonon and diffusion processes simultaneously. This is done by working through the whole cycle from polarisation data, through to the wavelength intensity and scattering functions, up the intermediate scattering function where physicallly significant properties can be deduced. Not only has this made current helium spin-echo experiments more convenient, but this has also brought forward future projects that were originally deemed too time-consuming to measure. Eventually the final goal is to capture the entire scattering function over all of $(\mb{K}, E)$-space using a two-dimensional continuous CS method. The authors hope that these advances will be quickly brought to the attention of the neutron scattering and X-ray communities.
\vspace{10pt}

\textbf{Acknowledgments:}
A. Jones acknowledges EPSRC grant EP/H023348/1, A. Tamt{\"o}gl acknowledges support by the FWF (project J3479-N20), I. Calvo-Almaz\'an acknowledges support from the Ram\'on Areces Institution and A. Hansen acknowledges support from  a Royal Society University Research Fellowship as well as EPSRC grant EP/L003457/1. The authors would like to thank W. Allison for many helpful discussions.

\bibliographystyle{unsrt}
\bibliography{biblio}

\begin{thebibliography}{10}

\bibitem{He3Apparatus}
A.~P. Jardine, H.~Hedgeland, G.~Alexandrowicz, W.~Allison, and J.~Ellis.
\newblock Helium-3 spin-echo: Principles and application to dynamics at
  surfaces.
\newblock {\em Prog. Surf. Sci.}, 84:323--379, 2009.

\bibitem{Bragg}
W.~L. Bragg.
\newblock The diffraction of short electromagnetic waves by a crystal.
\newblock {\em Proceedings of the Cambridge Philosophical Society}, 17:43–57,
  1913.

\bibitem{ohninereview}
A.~P. Jardine, G.~Alexandrowicz, H.~Hedgeland, W.~Allison, and J.~Ellis.
\newblock Studying the microscopic nature of diffusion with helium-3 spin-echo.
\newblock {\em Phys. Chem. Chem. Phys.}, 11(18):3355--3374, 2009.

\bibitem{oldphononpaper}
U.~Harten, J.~P. Tonnies, and Ch. Woll.
\newblock Helium time-of-flight spectroscopy of surface-phonon dispersion
  curves of the noble metals.
\newblock {\em Faraday Discuss. Gem. Soc.}, 80:137--149, 1985.

\bibitem{vanHove}
L.~van Hove.
\newblock Correlations in space and time and born approximation scattering in
  systems of interacting particles.
\newblock {\em Phys. Rev.}, 95(1):249, 1954.

\bibitem{CandesTao}
E.~J. Cand{\`e}s, J.~Romberg, and T.~Tao.
\newblock Robust uncertainty principles: exact signal reconstruction from
  highly incomplete frequency information.
\newblock {\em IEEE Trans. Inform. Theory}, 52(2):489–509, 2006.

\bibitem{Donoho}
D.~L. Donoho.
\newblock Compressed sensing.
\newblock {\em IEEE Trans. Inform. Theory}, 52(4):1289–1306, 2006.

\bibitem{CompressedMRI}
M.~Lustig, D.~L. Donoho, J.~M. Santos, and J.~M. Pauly.
\newblock Compressed sensing mri.
\newblock {\em IEEE Signal Processing Magazine}, 25(2):72--82, March 2008.

\bibitem{Unser}
M.~Guerquin-Kern, M.~H{\"{a}}berlin, K.P. Pruessmann, and M.~Unser.
\newblock A fast wavelet-based reconstruction method for magnetic resonance
  imaging.
\newblock {\em {IEEE} Transactions on Medical Imaging}, 30(9):1649--1660, 2011.

\bibitem{CompressedSpectroscopy}
D.~J. Holland, M.~J. Bostock, L.~F. Gladden, and D.~Nietlispach.
\newblock Fast multidimensional nmr spectroscopy using compressed sensing.
\newblock {\em Angewandte Chemie International Edition}, 50(29):6548–6551,
  2011.

\bibitem{RamenApplication}
D.~Galvis-Carreno, Y.~Meijia-Melgarejo, and H.~Arguello-Fuentes.
\newblock Efficient reconstruction of raman spectroscopy imaging based on
  compressive sensing.
\newblock {\em DYNA}, 2014.

\bibitem{MDApplication}
X.~Andrade, J.~N. Sanders, and A.~Aspuru-Guzik.
\newblock Application of compressed sensing to the simulation of atomic
  systems.
\newblock {\em PNAS}, 108(35):13928--13933, 2012.

\bibitem{ContinuousCompressed}
B.~Adcock and A.~Hansen.
\newblock Generalized sampling and infinite dimensional compressed sensing.
\newblock {\em Found. Comp. Math.}, 2015.

\bibitem{LeveledCompressed}
B.~Adcock, A.~C. Hansen, C.~Poon, and B.~Roman.
\newblock Breaking the coherence barrier: A new theory for compressed sensing.
\newblock {\em (Preprint)}, 2014.

\bibitem{AHBogdanTeshckeGenSamp}
B.~Adcock, A.~C. Hansen, B~Roman, and G~Teschke.
\newblock Generalized sampling: stable reconstructions, inverse problems and
  compressed sensing over the continuum.
\newblock {\em Adv. in Imag. and Electr. Phys.}, 182:187–279, 2014.

\bibitem{Roman}
B.~Roman, B.~Adcock, and A.~Hansen.
\newblock On asymptotic structure in compressed sensing.
\newblock {\em Preprint}, 2014.

\bibitem{WaveletSampling2}
B.~Adcock, A.~Hansen, and B.~Roman.
\newblock The quest for optimal sampling: Computationally efficient,
  structure-exploiting measurements for compressed sensing.
\newblock {\em Springer}, 2015.

\bibitem{Foundations}
B.~Roman, A.~Bastounis, B.~Adcock, and A.~Hansen.
\newblock On fundamentals of models and sampling in compressed sensing.
\newblock {\em Preprint}, 2015.

\bibitem{TransportAdsorbates}
J.~V. Barth.
\newblock Transport of adsorbates at metal surfaces: from thermal migration to
  hot precursors.
\newblock {\em Surfaces Science Reports}, 40:75--149, 2000.

\bibitem{ThermalCarbonNanotubes}
Amelia Barreiro, Riccardo Rurali, Eduardo, R.~Hernández, Joel Moser, Thomas
  Pichler, László Forró, and Adrian Bachtold.
\newblock Subnanometer motion of cargoes driven by thermal gradients along
  carbon nanotubes.
\newblock {\em Science}, 320:775--778, 2008.

\bibitem{AtomicWaterWheels}
Eunan J.~McEniry Daniel Dundas~and and Tchavdar~N. Todorov.
\newblock Current-driven atomic waterwheels.
\newblock {\em Nature Nanotechnology}, 4:99--102, 2009.

\bibitem{SurfacePhysicsIntro}
P.~Hofmann.
\newblock {\em Surface Physics: An Introduction}.
\newblock eBook 978-87-996090-1-7, 2013.

\bibitem{SolidStateIntro}
C.~Kittel.
\newblock {\em Introduction to Solid State Physics}.
\newblock John Wiley \& Sons Inc, 2005.

\bibitem{Benedek2010}
G~Benedek, M~Bernasconi, V~Chis, E~Chulkov, P~M Echenique, B~Hellsing, and
  J~Peter Toennies.
\newblock Theory of surface phonons at metal surfaces: recent advances.
\newblock {\em Journal of Physics: Condensed Matter}, 22(8):084020, 2010.

\bibitem{Kress1991}
Winfried Kress, de~Wette, and Frederik W.
\newblock {\em Surface Phonons}.
\newblock Number~21 in Springer Series in Surface Sciences. Springer Berlin
  Heidelberg, 1991.

\bibitem{Heid2003}
R.~Heid and K.~P. Bohnen.
\newblock Ab initio lattice dynamics of metal surfaces.
\newblock {\em Physics Reports}, 387(5–6):151--213, November 2003.

\bibitem{GilJardine2D}
G.~Alexandrowicz and A.~P. Jardine.
\newblock Helium spin-echo spectroscopy: studying surface dynamics with
  ultra-high-energy resolution.
\newblock {\em J. Phys.: Condens. Matter}, 19, 2007.

\bibitem{2DRecent}
E.~M. McIntosh, P.~R. Kole, M.~El-Batanouny, D.~M. Chisnall, J.~Ellis, , and
  W.~Allison.
\newblock Measurement of the phason dispersion of misfit dislocations on the
  au(111) surface.
\newblock {\em Phys. Rev. Lett.}, 110, 2013.

\bibitem{SPGL}
E.~Berg and M.~Friedlander.
\newblock Probing the pareto frontier for basis pursuit solutions.
\newblock {\em SIAM Journal of Scientific Computing}, 31(2):890--912, 2008.

\bibitem{Candes_PNAS}
V.~Studer, J.~Bobin, M.~Chahid, H.~Moussavi, E.~Cand{\`{e}}s, and M.~Dahan.
\newblock Compressive fluorescence microscopy for biological and hyperspectral
  imaging.
\newblock {\em Natl Acad Sci USA}, 109(26):1679--1687, 2011.

\bibitem{Mallat}
St{\'e}phane Mallat.
\newblock {\em A wavelet tour of signal processing}.
\newblock Elsevier/Academic Press, Amsterdam, third edition, 2009.

\bibitem{DaubechiesCPAM}
I.~Daubechies.
\newblock Orthonormal bases of compactly supported wavelets.
\newblock {\em Comm. Pure Appl. Math.}, 41(7):909---996, 1988.

\bibitem{Squires}
G.~L. Squires.
\newblock {\em Introduction to the Theory of Thermal Neutron Scattering}.
\newblock Cambirdge University Press, 1978.

\bibitem{QuasiNeutron}
M.~B{\'e}e.
\newblock {\em Quasielastic Neutron Scattering}.
\newblock CRC Press, 1988.

\end{thebibliography}

\pagebreak
\begin{center}
\textbf{\large Supporting Information: Continuous Compressed Sensing of Inelastic and Quasielastic Helium Atom Scattering Spectra}
\end{center}
\setcounter{equation}{0}
\setcounter{figure}{0}
\setcounter{table}{0}
\setcounter{page}{1}
\makeatletter
\renewcommand{\theequation}{S\arabic{equation}}
\renewcommand{\thefigure}{S\arabic{figure}}
\subsection{From Solenoid Currents to the Spin Phase}
The polarisation of the helium spin can be encoded by a phase quantity $\phi$. We assume that the relationship between the generated field strength $B$ in the solenoids and the current $I$ flowing through the coil is linear, i.e. $B=B_{\mrm{eff}} \cdot I$ for some constant $B_{\mrm{eff}}$. If the solenoid has length $L$ then the total accumulated phase has the form
\begin{equation} \label{Eq:AccumulatedPhase}
\phi= \frac{\gamma}{V}  B_{\mrm{eff}} I  ,
\end{equation}
where $V$ denotes the velocity of the He atom and $\gamma$ is the gyromagnetic ratio of the He atom. Therefore if $\phi_i$ denotes the accumulated phase in the first coil and $\phi_f$ the phase in the second then
\begin{equation} \label{Eq:TotalPhase}
\phi = \phi_i + \phi_f = \gamma  B_{\mrm{eff}}    \Bigg( \frac{ I_i}{V_i}+\frac{ I_f}{V_f} \Bigg),
\end{equation}
where we assume that the currents $I_i, I_f$ and velocities $V_i, V_f$ are different but the length $L$ and constant $B_{\mrm{eff}}$ is the same. We observe that the incoming velocity $V_i$ is related to the incoming wavelength via the de Broglie relation, i.e. $V_i=p_i m^{-1} = h (m \lambda_i)^{-1}$, where $\lambda_i$ denotes the wavelength of the beam. Consequently, \eqref{Eq:TotalPhase} becomes
\begin{equation} \label{Eq:WavelengthPhaseRelation}
 \phi=m \gamma B_{\mrm{eff}} h^{-1}  \big( I_i \lambda_i+ I_f \lambda_f \big).
\end{equation}
Now suppose that polarisation is rotated in the $xy$-plane and is initially polarised in the $x$-direction. Then, assuming that the analyser near the detector is also in the $x$-direction, the signal received has the form \cite{He3Apparatus}
\begin{equation} \label{Eq:TotalSignalWavelengthMessy}
\begin{aligned}
&P_{x,I}(I_i,I_f)= \langle \cos \phi \rangle_\rho
\\
&= \int \rho(\lambda_i,\lambda_f) \cos \Big(  m \gamma B_{\mrm{eff}} h^{-1}  \big( I_i \lambda_i+ I_f \lambda_f \big) \Big) \D \lambda_i \D \lambda_f.
\end{aligned}
\end{equation}
Here $\rho(\lambda_i, \lambda_f)$ denotes the \emph{Wavelength Intensity Function} which describes the distribution of helium atoms that reach the final polariser according to initial and final wavelengths. When combining the polarisation along the $x$ and $y$-direction after \eqref{Eq:TotalSignalWavelengthMessy} to a complex quantity and by introducing the scaled variables $(\kappa_i,\kappa_f)$ which have been defined in equation \ref{Eq:Current2Kappa} in the main text the polarisation becomes:
\begin{equation} \label{Eq:TotalSignalWavelength2piVerA}
\begin{aligned}
&P(\pmb{\kappa})  = \int \rho(\pmb{\lambda}) e^{2 \pi \ri \pmb{\kappa} \cdot \pmb{\lambda}} \, \D \pmb{\lambda}, \\
 \pmb{\lambda}&=(\lambda_i,\lambda_f) , \pmb{\kappa}=(\kappa_i,\kappa_f) \in \bbR^2.
\end{aligned}
\end{equation}
which is the two-dimensional Fourier transform introduced in equation \ref{Eq:TotalSignalWavelength2piVer} of the main text.

\section{Experimental Details}
\subsection{Sample Preparation}
The single crystals used in the study were discs with a diameter of 10 mm and a thickness of 1 mm. The crystals were mounted on the sample holder which can be heated using a radiative heating from a filament on the backside of the crystal or cooled down to 100 K using liquid nitrogen. The sample temperature was measured using a chromel-alumel thermocouple.

Prior to the measurements the surface was cleaned by several Ar$^{+}$ sputtering and annealing cycles. For the Au(111) surface this included cycles of sputtering with 0.5 kV Ar$^+$ ions, 5 \textmu A current for 15 minutes followed by annealing to 800 K (1 min). Ag(001) was typically sputtered with 0.8 kV Ar$^+$ ions, 8 \textmu A current for 20 minutes, followed again by annealing to 800 K (2 min). The base pressure in the scattering chamber was $< 3 \cdot 10^{-11}$ mbar and the surface quality was monitored using helium reflectivity measurements.

The Au(111) experiments were performed at a sample temperature of $T_S = 200$ K, at which the surface remained clean, with no measured decrease in the reflectivity, for a period of at least four hours. After this time any adsorbed contaminants were removed by flashing the crystal to 500 K before continuing measurements. The Ag(001) experiments were done with the crystal above room temperature where the surface remains clean for several days.

For the  deposition of cobalt phthalocyanine (CoPc, C$_{32}$H$_{16}$CoN$_8$) on Ag(001), a home-built Knudsen cell was used where a crucible filled with CoPc is resistively heated. The Knudsen cell was mounted in a separate dosing arm and was inserted into the scattering chamber for deposition of CoPc onto the clean Ag(001) surface. CoPc was deposited at a surface temperature of 350 K and a typical dosing pressure of $2 \cdot 10^{-9}$ mbar until the elastically scattered He signal had been attenuated by a factor of 2.

\subsection{Experimental Parameters}
The experimental parameters for the measurements presented in this study are summarised in table \ref{tab:ExperimentTable}.
\begin{table}[htbp]
\caption{Experimental parameters for the measured systems presented in this work.}
\begin{center}
\begin{tabular}{l|r|r}
\toprule
Measured system & Au(111) & CoPc/Ag(001)\\
\midrule
Sample temperature  & 200 K & 350 K \\
Incident He energy $E_i$ & 8.0 meV & 8.1 meV \\
Current range & [-4,4] A  & [0,10] A \\
Number of sampled points & 2048 & 101 \\
Incident angle $\theta_i$ & 19.325$^{\circ}$ & 24.2$^{\circ}$ \\
Momentum transfer $\Delta K$ & \multirow{2}{40pt}{\raggedleft $0.32\;\mbox{\AA}^{-1}$} & \multirow{2}{60pt}{\raggedleft $-0.22\;\mbox{\AA}^{-1}$} \\
(for elastic scattering) & & \\
\bottomrule
\end{tabular}
\end{center}
\label{tab:ExperimentTable}
\end{table}

\end{document}